\documentclass[12pt]{article}
\usepackage{epsfig}
\usepackage{amssymb}
\usepackage{rotating}
\usepackage{graphics}

\def\a{\alpha}
\def\b{\beta}

\def\d{\delta}

\def\g{\gamma}

\def\k{\kappa}

\def\o{\omega}

\def\r{\rho}
\def\s{\sigma}

\def\D{\Delta}

\def\G{\Gamma}

\def\O{\Omega}

\def\ve{\varepsilon}

\def\mt{\widetilde{m}_1}

\def\be{\begin{equation}}
\def\ee{\end{equation}}

\def\bea{\begin{eqnarray}}
\def\eea{\end{eqnarray}}

\def\pl#1#2#3{Phys.~Lett.~{\bf B {#1}} ({#2}) #3}
\def\np#1#2#3{Nucl.~Phys.~{\bf B {#1}} ({#2}) #3}
\def\prl#1#2#3{Phys.~Rev.~Lett.~{\bf #1} ({#2}) #3}
\def\pr#1#2#3{Phys.~Rev.~{\bf D {#1}} ({#2}) #3}

\begin{document}
\date{}

\title{
{\normalsize
\mbox{ }\hfill
\begin{minipage}{3cm}
MPP-2006-20
\end{minipage}}\\
\vspace{3mm}
\bf Leptogenesis beyond the limit
of hierarchical heavy neutrino masses}
\author{Steve Blanchet and Pasquale Di Bari
\\
{\it Max-Planck-Institut f\"{u}r Physik} \\ {\it (Werner-Heisenberg-Institut)} \\
{\it F\"{o}hringer Ring 6, 80805 M\"{u}nchen, Germany}}

\maketitle

\thispagestyle{empty}

\vspace{-4mm}
\centerline{\date{\today}}

\vspace{-2mm}
\begin{abstract}
\noindent
We calculate the baryon asymmetry of the Universe in thermal
leptogenesis beyond the usual lightest  right-handed (RH) neutrino dominated
scenario ($N_1$DS) and in particular beyond the hierarchical limit (HL),
$M_1\ll M_2\ll M_3$, for the RH neutrino mass spectrum.
After providing some orientation among the
large variety of models, we first revisit the central role of the
$N_1$DS, with new insights on the dynamics of the asymmetry
generation and then discuss
the main routes departing from it, focusing on models beyond the HL.
We study in detail two examples of `strong-strong' wash-out scenarios:
one with `maximal phase' and the limit of very large $M_3$,
studying the effects arising when $\d_2\equiv (M_2-M_1)/M_1$ is small.
We extend analytical methods already applied to the $N_1$DS
showing, for example, that, in the degenerate limit (DL),
the efficiency factors of the RH neutrinos become
equal with the single decay parameter
replaced by the sum. Both cases disprove
the misconception that close RH neutrino masses necessarily lead
to a final asymmetry enhancement and to a relaxation of the lower
bounds on $M_1$ and on the initial temperature of
the radiation-dominated expansion. We also explain why leptogenesis
tends to favor normal hierarchy compared to inverted hierarchy
for the left-handed neutrino masses.

\end{abstract}

\newpage
\section{Introduction}

With the discovery of neutrino masses in neutrino mixing
experiments, leptogenesis \cite{fy} has become one of the most
attractive explanations of the matter-antimatter asymmetry of the
Universe. Indeed, leptogenesis is the direct cosmological
consequence of the see-saw mechanism \cite{seesaw}, the most
elegant way to understand neutrino masses and their lightness
compared to all other known massive fermions. Adding to the
Standard Model Lagrangian three RH neutrinos with Yukawa coupling
matrix $h$ and Majorana mass matrix $M$, a neutrino Dirac mass
matrix $m_D=h\,v$ is generated, after electroweak symmetry
breaking, by the vev $v$ of the Higgs boson. For $M\gg m_D$, the
neutrino mass spectrum splits into 3 heavy Majorana states $N_1$,
$N_2$ and $N_3$ with masses $M_1\leq M_2\leq M_3$, which almost
coincide with the eigenvalues of $M$, and 3 light Majorana states
with masses $m_1\leq m_2 \leq m_3$ corresponding to the
eigenvalues of the neutrino mass matrix given by the see-saw
formula,
\be\label{seesaw}
m_{\nu}=-m_D\,{1\over M}\,m_D^T \, .
\ee
Neutrino mixing experiments measure two mass-squared differences.
In normal (inverted) neutrino schemes, one has
\be
m^2_3-m^2_2= \D m^2_{\rm atm}\,\,(\D m_{\rm sol}^2) \,\, ,
\ee
\be
m^2_2-m^2_1= \D m^2_{\rm sol}
\,\,(\D m_{\rm atm}^2-\D m^2_{\rm sol}) \, .
\ee
For $m_1\gg m_{\rm atm}\equiv
\sqrt{\D m_{\rm atm}^2+\D m_{\rm sol}^2}
\simeq 0.05\,{\rm eV}$, one has a quasi-degenerate
spectrum with $m_1\simeq m_2 \simeq m_3$, whereas for
$m_1\ll m_{\rm sol}\equiv \sqrt{\D m^2_{\rm sol}}\simeq 0.009\,{\rm eV}$
one has a fully hierarchical (normal or inverted) spectrum.

 A lepton asymmetry can be generated from the decays of the
heavy neutrinos into leptons and Higgs bosons and
partly converted into a baryon asymmetry by the sphaleron
($B-L$ conserving) processes at temperatures higher
than about $100\,{\rm GeV}$.
The asymmetry produced by each $N_i$ decay is given by the $C\!P$
asymmetry parameter $\ve_i$
\be
\ve_i\equiv -\,{\G_i-\bar{\G}_i\over \G_i+\bar{\G}_i} \, ,
\ee
where $\G_{\rm i}$ is the decay rate into leptons and
$\bar{\G}_{\rm i}$ the one into anti-leptons.
 For each $N_i$ one introduces the decay parameter $K_i$,
defined as the ratio of the total decay width to the
expansion rate at $T=M_i$,
\be\label{decpar}
K_i\equiv {\widetilde{\G}_{i}\over H(T=M_i)} \, .
\ee
This is the  key quantity for the thermodynamical
description of the decays of heavy particles in the early Universe
\cite{kt}.
In leptogenesis it can be conveniently expressed in terms
of the {\em effective neutrino mass} $\widetilde{m}_i\equiv
{(m_D^{\dagger}\,m_D)_{ii}/M_i}$,
such that $K_i=\widetilde{m}_{i}/m_{\star}$, where
$m_{\star} \simeq 1.08\times 10^{-3}\,{\rm eV}$
is the {\em equilibrium neutrino mass}.
Besides decays, there are other processes,
especially inverse decays, that are
relevant not only for producing the RH neutrinos
but also for washing-out part of the asymmetry produced from decays.
The effect of production and  wash-out are simultaneously
accounted for by the efficiency factors
$\k_i$ associated to the production of the asymmetry from each $N_i$,
such that the final $B-L$ asymmetry can be expressed as the sum of
three contributions
\be\label{NBmL}
N_{B-L}^{\rm f} = \,\sum_i\,\ve_i\,\k_i^{\rm f} \, .
\ee
The baryon-to-photon number ratio at the recombination time
can then be calculated as
\be\label{general}
\eta_B=\,a_{\rm sph}\,{N_{B-L}^{\rm f}\over N_{\gamma}^{\rm rec}}
\simeq 0.96\times 10^{-2}\,\sum_i\,\ve_i\,\k_i^{\rm f} \, ,
\ee
where $a_{\rm sph}\simeq 1/3$ is the sphaleron
conversion coefficient. Here we assume a standard thermal
history and indicate with $N_X$ any particle number or asymmetry $X$
calculated in a portion of comoving volume containing one heavy neutrino
in ultra-relativistic thermal equilibrium, so that
$N_{\gamma}^{\rm rec}\simeq 37$. The efficiency factors
$\k_i^{\rm f}\rightarrow 1$ in the limit of an initial
ultra-relativistic thermal $N_i$ abundance and null wash-out.

A great simplification occurs in the HL
\footnote{Throughout the paper we make use of
the following acronyms: HL=hierarchical limit,
DL=degenerate limit, $N_i$DS=$N_i$-dominated scenario.}
($M_1\ll M_2 \ll M_3$).
In this case one typically (but not necessarily!)
obtains what can be called the $N_1$DS,
where both the wash-out from the two heavier RH neutrinos and
the asymmetry produced by their decays can be neglected
and the expression (\ref{NBmL}) reduces
to  $(N_{B-L}^{\rm f})_{\rm HL}\simeq \,\k_1^{\rm f}\,\ve_1$.
The HL is quite a natural assumption for hierarchical light neutrinos.

The effective neutrino mass
$\mt$ can be expressed as a linear combination of the
light neutrino masses with positive coefficients whose sum
cannot be smaller than unity. For this reason,
the experimental findings $m_{\rm sol},m_{\rm atm}\gg m_{\star}$
typically force $K_1$ to lie
in the range
${\cal O}(K_{\rm sol}\simeq 9)\lesssim K_1 \lesssim
{\cal O}(K_{\rm atm}\simeq 50)$,
where $K_{\rm sol}\equiv {m_{\rm sol}/m_{\star}}$
and
\mbox{$K_{\rm atm}\equiv {m_{\rm atm}/m_{\star}}$}, i.e.
in the strong wash-out regime ($K_1\gg 1$), while
the weak wash-out regime ($K_1\lesssim 1$) is possible
for a particular class of neutrino mass models.

The efficiency factor $\k_1^{\rm f}$ is approximately given by
the {\em number of $N_1$ that decay out-of-equilibrium}.
In the strong wash-out regime this is unambiguously specified
by the thermal equilibrium abundance at the time when the
inverse decays get frozen, at a well-defined temperature
$T_B\ll M_1$ when the $N_1$'s are non-relativistic \cite{annals}.
One has then to require that the initial temperature
of the Universe is larger than $\sim T_B$,
the key assumption for thermal leptogenesis.
Therefore, in the strong wash-out regime
only a small fraction of $N_1$, compared to an
initial ultra-relativistic
thermal abundance, decays out-of-equilibrium.
This results in small values for
$\k_1^{\rm f}\sim 10^{-3}\div 10^{-2}$,
but still large enough to allow for successful leptogenesis in
quite a large region of parameter space.
On the other hand,
the positive by-product of  the strong wash-out regime is that
the final asymmetry does not depend on the initial conditions.

A second important simplification occurring in the HL
is that the $C\!P$ asymmetry $\ve_1$, like $\k_1^{\rm f}$,
depends only on a limited set of see-saw parameters and,
quite remarkably, it turns out that there is an upper bound on $\ve_1$
proportional to $M_1$ \cite{asaka,di02}.
In the strong wash-out regime for $K_1\gtrsim 5$,
this gives rise to a lower bound on the lightest RH neutrino mass
$M_1\gtrsim 5\times 10^{9}\,{\rm GeV}$ \cite{cmb,di02,geometry},
also implying a lower bound on the initial temperature of the
radiation-dominated expansion
$T_{\rm in}\gtrsim 2\times 10^{9}\,{\rm GeV}$~\cite{annals,geometry}
\footnote{In the MSSM these values become respectively
$M_1\gtrsim 2 \times 10^{9}\,{\rm GeV}$ and
$T_{\rm in}\gtrsim 10^{9}\,{\rm GeV}$ \cite{giudice,proc,geometry}.},
identifiable with the reheating temperature within inflation \cite{gkr}.
 In the $N_1$DS, for quasi-degenerate neutrinos,
  the combined effect \cite{bound1}
 of the additional wash-out from $\Delta L=2$ processes,
 which depends on the combination $M_1\,\sum\,m_i^2$, together with a
 $C\!P$ asymmetry suppression \cite{di02,bound2}, gives rise to a stringent
 upper bound on the absolute neutrino mass scale
 \mbox{$m_1\leq 0.1\,{\rm eV}$} \cite{bound1,bound2,giudice,annals}.

  A weak version of the $N_1$DS, for $K_1\lesssim 1$,
encounters two serious difficulties.
  The first is the strong dependence on the
initial conditions, preventing the model from being self-contained.
The second is that, for $\mt \rightarrow m_1$, the $C\!P$
 asymmetry vanishes and $\mt$ has to be fine-tuned
 to have successful leptogenesis.
A more  appealing possibility is then represented by the $N_2$DS \cite{geometry},
where the asymmetry is mostly generated from the decays of $N_2$,
circumventing both problems.

Another key motivation to study models beyond the HL is to
allow RH neutrino masses to be arbitrarily close.
This possibility has been considered in many works \cite{ke,many,branco}.
In \cite{geometry} an analytical condition for
the validity of a calculation of the efficiency
factor $\k_1^{\rm f}$ in the HL was found.
It was noticed that one may neglect
the effect of the two heavier neutrinos if
$\d_2\equiv {(M_2-M_1)/ M_1}\gtrsim 1.5\div 5$,
the exact value depending on $K_1$ and $K_2$.
The validity of the HL in the calculation of the
$C\!P$ asymmetry is more involved but a similar condition
holds in most cases. There is one possibility, discussed in the
case C of Section 4, where the condition on $\d_2$ to recover
the HL becomes more stringent. For quasi-degenerate neutrinos
this also provides a way to evade the upper bound on neutrino masses
\cite{hambye}.

In this paper we perform a general calculation
of the final asymmetry beyond the HL, extending  useful
analytical methods described in \cite{annals} within the HL.
A discussion of flavor effects \cite{bcst,newflavor1,newflavor2}
is deferred to a forthcoming paper, since they are somewhat
complementary to the issues addressed here. However, it is worthwhile
to mention that, accounting for these effects,
the upper bound on the absolute neutrino mass scale does not hold
\cite{newflavor2}.

The main difficulty of such a general calculation is
the great model dependence. Therefore, in Section 2
our first step is a description of a general
way to parameterize and classify models.
In Section 3 we revisit the $N_1$DS,
providing several new interesting analytical insights on the
dynamics of the asymmetry generation. In Section 4
we describe the main routes to go beyond the $N_1$DS,
including models beyond the HL that we study in detail in Section 5.
Here we focus on {\em strong-strong} wash-out scenarios,
where both $K_1$ and $K_2 \gtrsim 5$ and $M_3\gg M_1,M_2$,
but with arbitrary $M_1$ and $M_2$.
We show how the production and the wash-out
from each RH neutrino interfere with each other, calculating
the efficiency factors and giving exact conditions for the HL to be recovered.
Then we calculate the lower bounds on $M_1$ and $T_{\rm in}$
first in a model where the asymmetry is maximal in the HL, but
insensitive to a $C\!P$ asymmetry enhancement beyond the HL,
then in a model that received recently great attention, where
$M_3\gg 10^{14}\,{\rm GeV} \gg M_2,M_1$ \cite{fgy}.
We also explain why leptogenesis favors normal hierarchy over
an inverted one. We summarize our conclusions in Section 6.

\section{Getting oriented among leptogenesis scenarios}

Let us describe how to calculate the final asymmetry
for a general RH neutrino spectrum. In a one-flavor approximation,
the set of kinetic equations can be written as \cite{luty,ke,many}
\begin{eqnarray}
{dN_{N_i}\over dz} & = & -(D_i+S_i)\,(N_{N_i}-N_{N_i}^{\rm eq}) \;,
\hspace{10mm} i=1,2,3 \label{dlg1} \\
{dN_{B-L}\over dz} & = &
\sum_{i=1}^3\,\varepsilon_i\,D_i\,(N_{N_i}-N_{N_i}^{\rm eq})-N_{B-L}\,W \;,
\label{dlg2}
\end{eqnarray}
where $z\equiv M_1/T$. Defining $x_i\equiv M_i^2/M_1^2$
and $z_i\equiv z\,\sqrt{x_i}$, the {\em decay factors} are given by
\be
D_i \equiv {\G_{D,i}\over H\,z}=K_i\,x_i\,z\,
\left\langle {1\over\gamma_i} \right\rangle   \, ,
\ee
where $H$ is the expansion rate. The total decay rates,
$\G_{D,i}\equiv \G_i+\bar{\G}_i$,
are the product of the decay widths times the
thermally averaged dilation factors
$\langle 1/\gamma_i\rangle$, given by the ratio
${\cal K}_1(z_i)/ {\cal K}_2(z_i)$ of the modified
Bessel functions.
The equilibrium abundance and its rate are also expressed through
the modified Bessel functions,
\be
N_{N_i}^{\rm eq}(z_i)= {1\over 2}\,z_i^2\,{\cal K}_2 (z_i) \;\; ,
\hspace{10mm}
{dN_{N_i}^{\rm eq}\over dz_i} =
-{1\over 2}\,z_i^2\,{\cal K}_1 (z_i) \, .
\ee
The RH neutrinos can be produced by inverse decays and scatterings.
Nevertheless, in the relevant strong wash-out regime,
the inverse decays alone are already sufficient to make
the RH neutrino abundance
reach its thermal equilibrium value prior to their decays.
Therefore,
the details of the RH neutrino production do not affect the final
asymmetry and theoretical uncertainties are consequently
greatly reduced. This is one of the nice features of the
strong wash-out regime on which we will focus and
for this reason the scattering terms $S_i$ will
play no role.
The wash-out factor $W$ can be written as the sum
of two contributions \cite{cmb},
\be\label{W}
W=\sum_i\,W_i(K_i)+\D\, W(M_1\,\bar{m}^2) \, .
\ee
The second term arises from the non-resonant $\D L=2$ processes
and gives typically a non-negligible contribution only in the non
relativistic limit for $z\gg 1$ \cite{luty,cmb,annals}.
For hierarchical light neutrinos it can be safely
neglected for reasonable values $M_1\ll 10^{14}\,{\rm GeV}$.
In the strong wash-out regime, the first term
is dominated by inverse decays \cite{giudice,annals},
where the resonant $\D L=2$ contribution has to be
properly subtracted \cite{dolgov,giudice}, so that
\footnote{In the following we will imply
this subtraction when referring to the `wash-out
from inverse decays'.}
\be\label{WID}
W_i(z)\simeq W_i^{\rm ID}(z) =
{1\over 4}\,K_i\,\sqrt{x_i}\,{\cal K}_1(z_i)\,z_i^3 \, .
\ee
Let us indicate with $N_{B-L}^{\rm in}$ a possible pre-existing
asymmetry at $T_{\rm in}$.
The final asymmetry can then be written in
an integral form  \cite{kt,annals} ,
\be\label{integral}
N_{B-L}^{\rm f}=N_{B-L}^{\rm in}\,e^{-\sum_i \,\int\,dz'\,W_{i}(z')}+
\sum_i\,\ve_i\,\k_{i}^{\rm f} \, ,
\ee
with the efficiency factors $\k_i^{\rm f}$ given by
\begin{equation}\label{ki}
\k_i^{\rm f}= - \int_{z_{\rm in}}^{\infty}\, dz'\
{dN_{N_i}\over dz'}\
e^{-\sum_i\,\int_{z'}^{z}\ dz''\, W_{i}(z'')}\;,
\end{equation}
where we defined $z_{\rm in}\equiv M_1/T_{\rm in}$.
Notice that, in general, each efficiency factor depends on all decay parameters,
i.e. $\k_{i}^{\rm f}=\k_i^{\rm f}(K_1,K_2,K_3)$.

If the mass differences satisfy the condition for the
applicability of perturbation theory, $|M_j-M_i|/M_i \gg {\rm
max}[(h^{\dagger}\,h)_{ij}]/(16\,\pi^2)$ with $j\neq i$
\cite{alicia}, then a perturbative calculation from the
interference of tree level with one loop self-energy and vertex
diagrams gives \cite{CPas}
\be\label{CPas} \ve_i =\, {3\over 16\pi}\, \sum_{j\neq i}\,
{{\rm Im}\,\left[(h^{\dagger}\,h)^2_{ij}\right] \over
(h^{\dagger}\,h)_{ii}} \,{\xi(x_j/x_i)\over \sqrt{x_j/x_i}}\, \ee
where the function $\xi(x)$, shown in Fig.~1, is defined as
\cite{bound2} \be\label{xi} \xi(x)= {2\over 3}\,x\,
\left[(1+x)\,\ln\left({1+x\over x}\right)-{2-x\over 1-x}\right] \,
.
\ee
\begin{figure}
\centerline{\psfig{file=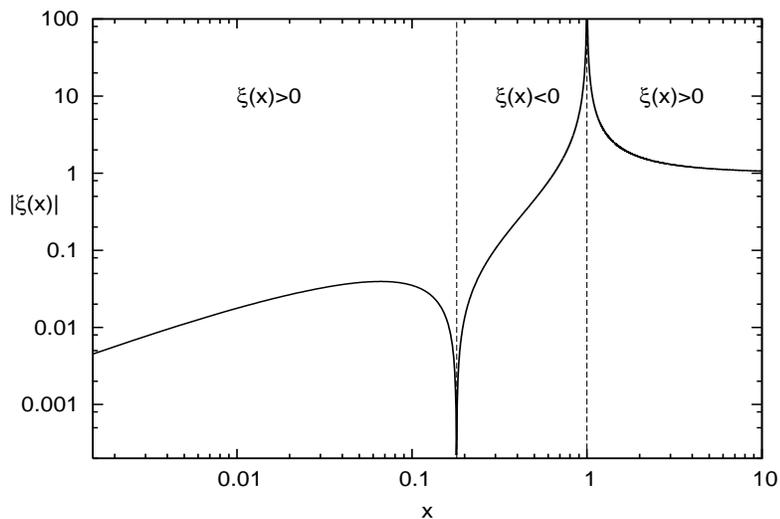,height=7cm,width=11cm}}
\caption{The function $\xi(x)$ defined in the Eq. (\ref{xi}).}
\end{figure}
Working in a basis where both
the Majorana mass and the light neutrino mass matrix are diagonal,
$D_M\equiv {\rm diag}(M_1,M_2,M_3)$ and $D_m\equiv {\rm
diag}(m_1,m_2,m_3)= -U^{\dagger}\,m_{\nu}\,U^{\star}$, from the
see-saw formula  one can obtain a useful parametrization of the
Dirac mass matrix in terms of the orthogonal complex matrix $\O$
\cite{casas} \footnote{Compared to the $R$ matrix in \cite{casas},
one has the simple relation $\O=R^{\dagger}$.},
\be\label{CI}
m_D=U\,\sqrt{D_m}\,\O\,\sqrt{D_M} \, .
\ee
The unitary matrix $U$ can be identified with the PMNS
matrix in a basis where the charged lepton mass matrix is also diagonal.
The following parametrization of the $\O$ matrix
proves to be particularly useful in leptogenesis \cite{geometry}
\be\label{first}
\O(\O_{21},\O_{31},\O_{22})=
\left(
\begin{array}{ccc}
 \sqrt{1-\O_{21}^2-\O_{31}^2} & \O_{12} &  \pm \sqrt{\O_{21}^2+\O_{31}^2-\O^2_{12}} \\
 \O_{21} & \O_{22} & - \sqrt{1-\O_{22}^2-\O_{21}^2} \\
 \O_{31} & \sqrt{1-\O_{22}^2-\O_{12}^2} & \sqrt{\O_{22}^2+\O^2_{12}-\O_{31}^2)}
\end{array}
\right) \, ,
\ee
where $\O_{12}$ can be expressed as a function of $(\O_{21},\O_{31},\O_{22})$
imposing, for example, $\sum_j\,\O_{j1}\,\O_{j2}=0$.
From this general form one can obtain, for particular choices
of the parameters, three elementary complex rotations that exhibit
peculiar properties. In terms of these rotations
an alternative parametrization of the $\O$ matrix is
\be\label{second}
\O({\o}_{21},{\o}_{31},{\o}_{22})
=R_{12}(\o_{21})\,\,
 R_{13}(\o_{31})\,\,
 R_{23}(\o_{22})\,\, ,
\ee
where
\be\label{R}
\mbox{\tiny $
R_{12}=
\left(
\begin{array}{ccc}
 \sqrt{1-{\o}^2_{21}}  &  -{\o}_{21}          & 0 \\
            {\o}_{21}  & \sqrt{1-{\o}^2_{21}} & 0 \\
  0 & 0 & 1
\end{array}
\right) \,
         \,\, , \,\,
R_{13}=
\left(
\begin{array}{ccc}
 \sqrt{1-{\o}^2_{31}}  & 0 &  - {\o}_{31} \\
    0 & 1 & 0 \\
  {\o}_{31} & 0 & \sqrt{1-{\o}^2_{31}}
\end{array}
\right)  \,\, ,
\,\,
R_{23}=
\left(
\begin{array}{ccc}
  1  &  0   & 0   \\
  0  & {\o}_{22} & -\sqrt{1-{\o}^2_{22}} \\
  0 &  \sqrt{1-{\o}^2_{22}} & {\o}_{22}
\end{array}
\right)$ \, .
}
\ee
This parametrization for an orthogonal complex matrix
corresponds to the transposed form of the
CKM  matrix in the quark sector or of the
PMNS unitary matrix in neutrino mixing, with the difference
that here one has complex rotations instead of real ones.
There are straightforward relations between  the parameters
$\O_{ij}$ and the $\o_{ij}$'s:
\be\label{rel}
\O_{21}  =  {\o}_{21}\,\sqrt{1-{\o}_{31}^2}\, , \hspace{3mm}
\O_{31}  =  {\o}_{31} \, , \hspace{3mm}
\O_{22}  = {\o}_{22}\,\sqrt{1-{\o}_{21}^2}-  \,
              {\o}_{21}\,{\o}_{31}\,\sqrt{1-{\o}_{22}^2} \, .
\ee
The two parameterizations are interchangeable and it can
be more convenient to use one or the other depending on the context.
The parametrization Eq.~(\ref{second}) is particularly useful
to understand the general structure of
different models occurring in thermal leptogenesis.
\begin{itemize}
\item For $\O=R_{13}$ one has $\ve_2=0$, while $\ve_1$ is maximal if
\cite{geometry} $ m_3\,{{\rm Re}(\o^2_{31})/ |\o_{31}^2|}=
           m_1\,{[1-{\rm Re}(\o^2_{31})]/ |1-\o_{31}^2|}$.
In the HL ($M_2\gg M_1$) one obtains the $N_1$DS.
\item
For $\O=R_{23}$ one has $\ve_1=0$, while $\ve_2$
is maximal if
$m_3\,{{\rm Re}(\o^2_{22})/|\o_{22}^2|}=
           m_2\,{[1-{\rm Re}(\o^2_{22})]/ |1-\o_{22}^2|}$.
At the same time, one has $\mt=m_1$, so that the wash-out from
$N_1$ can be neglected if $m_1$ is small enough.
Therefore, in the HL and for hierarchical light neutrinos,
one obtains the $N_2$DS \cite{geometry}, as discussed in Section 4.
\item If $\O=R_{12}$, then $\ve_1$ undergoes
a phase suppression compared to its maximal value but
$|\ve_2|\propto (M_1/M_2)\,|\ve_1|$. This implies that in the
HL one again recovers the $N_1$DS \cite{geometry}.
On the other hand
if $M_1\simeq M_2$ both $C\!P$ asymmetries can play a role.
\end{itemize}
Notice that $N_2$DS requires a more special $\O$ form
than the $N_1$DS, but on the other hand there is no lower bound
on $M_1$ as in the $N_1$DS. Therefore, it represents an interesting
alternative \cite{geometry}. On the other hand,
there cannot be a $N_3$DS with only 3 RH neutrinos. The
reason is that $\ve_3\rightarrow 0$ in the HL for any $\O$.
This can be understood more generally if one observes that
the $C\!P$ asymmetry of a decaying particle vanishes
in the limit where all particles in the propagators are massless. This also
explains why it is more special to have $|\ve_2|\gg |\ve_1|$
than the opposite: in the first case one must necessarily have
$N_3$ running in the propagator in order to have $\ve_2\neq 0$
and this happens for $\O=R_{23}$,
while in order to have $\ve_1\neq 0$ one can have
either $N_2$ for $\O=R_{12}$, or $N_3$ for $\O=R_{13}$,
or both.


\section{Revisiting the $\mathbf{N_1}$DS}

Let us now revisit shortly some of the results holding
in the $N_1$DS, with some new interesting insights
on the dynamics of the asymmetry generation. This is also necessary
to introduce quantities and notations that will be used or extended
in Section 5 to go beyond the HL.

The general expression
for the final asymmetry Eq.~(\ref{general}) reduces to
$\eta_B\simeq 10^{-2}\,\ve_1\,\k_1^{\rm f}$,
where $\k_1^{\rm f}$ can be calculated solving a
system of just two kinetic equations \cite{ke,cmb,giudice},
%
obtained from the general set (cf.~(\ref{dlg1}) and (\ref{dlg2}))
neglecting the asymmetry generation and the wash-out terms
from the two heavier RH neutrinos.

For $M_1 \ll 10^{14}\,{\rm GeV}\,(m_{\rm atm}^2/\sum\,m_i^2)$,
the term $\D W(z)$ in the wash-out (cf.~(\ref{W})) is negligible
and the solutions depend just on $K_1$,
since this is the only parameter in the equations.
They can be worked out in an integral form \cite{kt}
and for the $B-L$ asymmetry one obtains
a special case of the more general Eq. (\ref{integral}),
\be\label{NBmLf}
N_{B-L}(z;\bar{z})=\bar{N}_{B-L}\,e^{-\int_{\bar{z}}^{z}\,dz'\,W_{1}(z')}
+\ve_1\,\k_1(z;\bar{z}) \, ,
\ee
where now a possible asymmetry produced from the two heavier
RH neutrinos and frozen at $\bar{z}\geq z_{\rm in}$ is included in
$\bar{N}_{B-L}$.
The efficiency factor $\k_1(z;\bar{z})$ can be expressed through a Laplace integral,
\be\label{k1}
\k_1(z;\bar{z})=
-\int_{\bar{z}}^{z}\,dz'\,{dN_{N_1}\over dz'}\,
\,e^{-\int_{z'}^{\infty}\,dz''\,W_{1}(z'')}
=\int_{\bar{z}}^{z}\,dz'\,e^{-\psi(z',z)} \, .
\ee
In the strong wash-out regime, 
using the approximation $dN_{N_1}/dz'\simeq dN_{N_1}^{\rm eq}/dz'$
and $W_1(z')\simeq W_1^{\rm ID}(z')$ (cf.~(\ref{WID})), one finds
that for $z\rightarrow\infty$ the final value is given by \cite{annals}
\be\label{k}
\k_{1}^{\rm f}(K_1)\simeq \k(K_1) \equiv {2\over K_1\,z_B(K_1)}\,
\left(1-e^{-{K_1\,z_B(K_1)\over 2}}\right)  \, ,
\ee
if $\bar{z}\lesssim z_B-2$. The value $z'=z_B(K_1)$ is where the
quantity $\psi(z',\infty)$ has a minimum and the integral
in the Eq. (\ref{k1}) receives a dominant contribution from a restricted
$z'$-interval centered around it. In the strong wash-out regime
it is well reproduced by
\be\label{zB}
z_B(K_1) \simeq 2+4\,{K_1}^{0.13}\,e^{-{2.5\over K_1}} \, .
\ee
Figures 2 and 3 show,
for an initial thermal abundance,
the dynamics of the asymmetry generation,
comparing one example of  weak wash-out
with one example of strong wash-out
\footnote{Corresponding animations can be found at
http://wwwth.mppmu.mpg.de/members/dibari.}.
In the top panels we show
the function $d\k_1/dz'\equiv e^{-\psi(z',z)}$, defined for
$z'\leq z$, for different values of $z$.
The difference between the two cases is striking.
\begin{figure}[!h]
\vspace{40mm}
\begin{center}
\begin{picture}(100,100)
\put(-100,0){\includegraphics{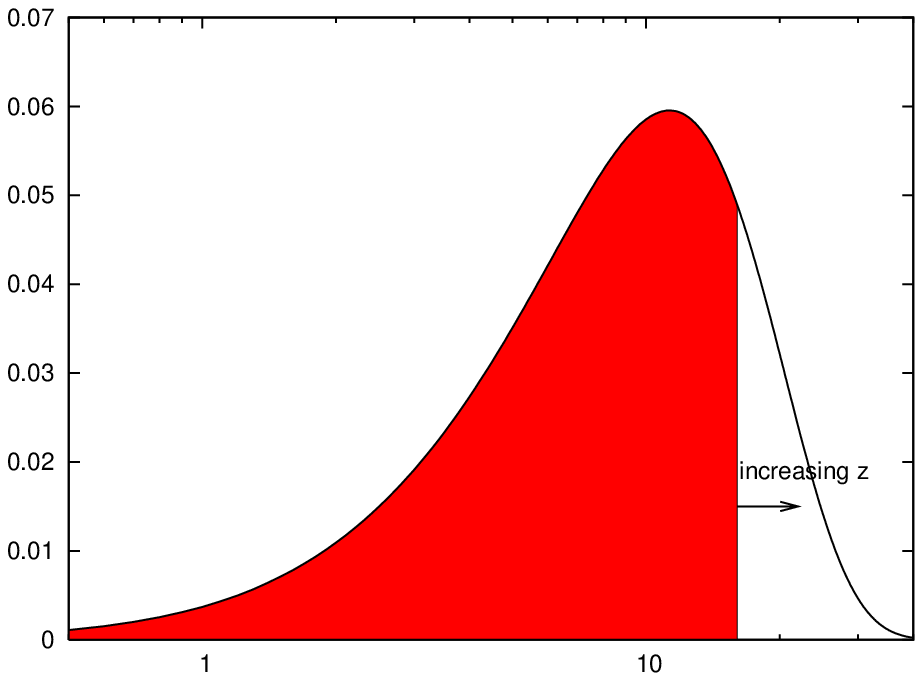}}
\put(-50,160){$K_1=10^{-2}$}
\put(95,200){$z_{\rm max}^{\rm weak}\simeq 12$}
\put(101,172){\Huge $\uparrow$}
\put(-35,115){$\left\arrowvert \frac{dN_{N_1}}{dz'}\right| \simeq
\left\arrowvert\frac{dN_{N_1}^{\rm weak}}{dz'} \right|$}
\put(15,80){\Huge $\searrow$}
\put(50,-10){$z'$}
\put(-132,100)
{\large $\left.\frac{d\kappa_1}{dz'}\right|_{z' \leq z}$}
\end{picture}
\end{center}
\vspace{28mm}
\begin{center}
\begin{picture}(100,100)
\put(-100,0){\includegraphics{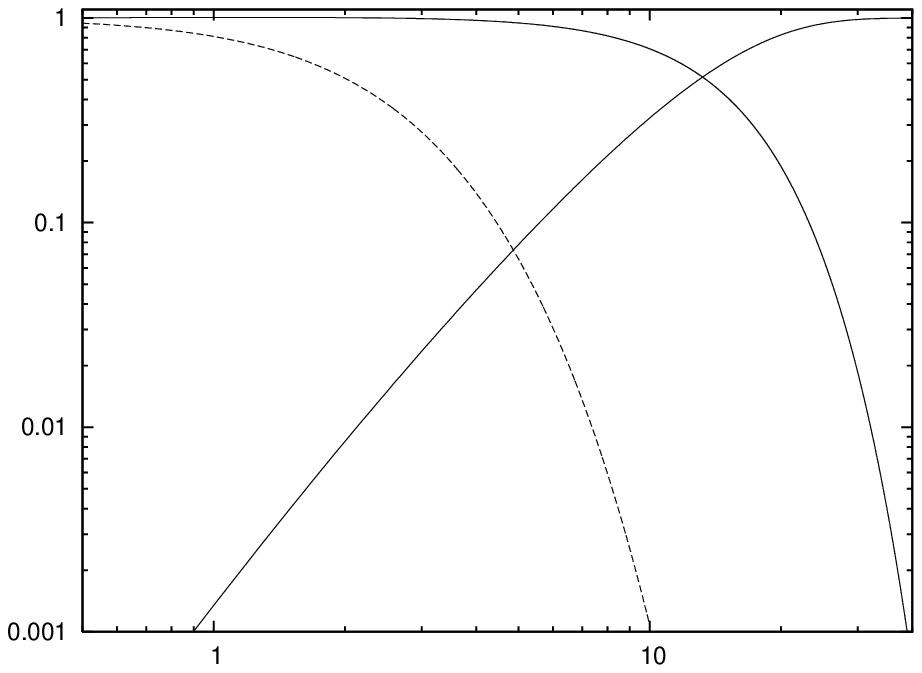}}
\put(-50,160){$K_1=10^{-2}$}
\put(50,-10){$z=\frac{M_1}{T}$}
\put(-25,50){$\kappa_1$}
\put(77,100){$N_{N_1}^{\rm eq}$}
\put(127,120){$N_{N_1}$}
\end{picture}
\end{center}
\caption{Dynamics in the weak wash-out regime for
initial thermal abundance ($N_{N_1}^{\rm in}=1$).
Top panel: rates. Bottom panel: efficiency factor $\k_1$ and $N_1$-abundance.
The maximum of the asymmetry production rate occurs at
$z'\simeq z_{\rm max}^{\rm weak}=1/\sqrt{K_1}+15/8\simeq 12$.}
\end{figure}
\begin{figure}[!h]
\vspace{5cm}
\begin{center}
\begin{picture}(100,100)
\put(-100,0){\includegraphics{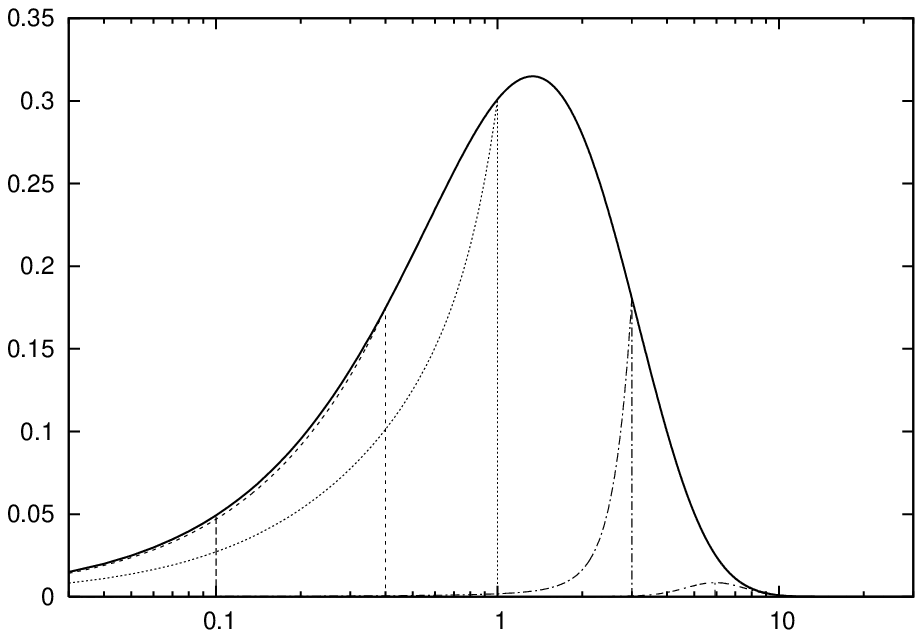}}
\put(-50,160){$K_1=10$}
\put(50,-10){$z'$}
\put(50,80){$z=1$}
\put(63,168){\Large $\uparrow$}
\put(50, 185){$z_B^{\rm eq}(0)\simeq 1.33$}
\put(-40,20){$z=0.1$}
\put(10,35){$z=0.4$}
\put(80,45){$z=3$}
\put(115,5){$z_B$}
\put(120,20){$\swarrow$}
\put(130,30){$z\gg z_B$}
\put(-50,135){$\left\arrowvert \frac{dN_{N_1}}{dz'}\right| \simeq
\left\arrowvert\frac{dN_{N_1}^{\rm eq}}{dz'} \right|$}
\put(3,105){\Huge $\searrow$}
\put(-132,100)
{\large $\left.\frac{d\kappa_1}{dz'}\right|_{z'\leq z}$}
\end{picture}
\end{center}
\vspace{30mm}
\begin{center}
\begin{picture}(100,100)
\put(-100,0){\includegraphics{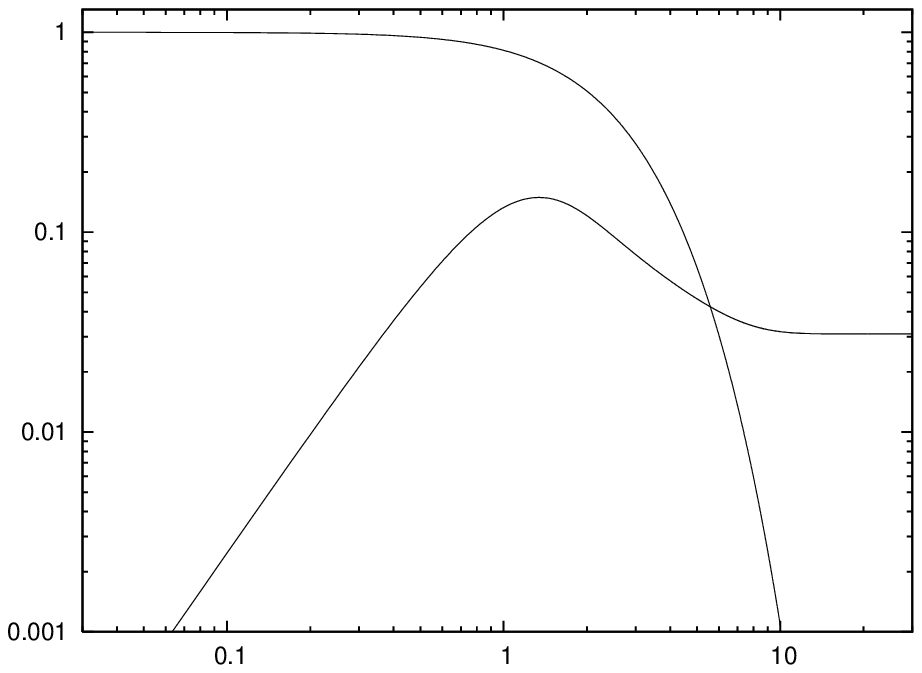}}
\put(-50,160){$K_1=10$}
\put(2,100){$\kappa_1$}
\put(50,-10){$z=\frac{M_1}{T}$}
\put(105,150){$N_{N_1}^{\rm eq} \simeq N_{N_1}$}
\end{picture}
\end{center}
\caption{Dynamics in the strong wash-out regime.
Top panel: rates. Bottom panel: efficiency factor $\k_1$ and $N_1$-abundance.
The maximum of the final asymmetry production rate
occurs at $z'\simeq z_B$.}
\end{figure}

In the weak wash-out
regime, each decay contributes to the final asymmetry for any value
of $z'$ when the asymmetry is produced.
In the strong wash-out regime, all asymmetry produced at $z'\lesssim z_B-2$
is efficiently washed-out by inverse decays, such that only decays
occurring at $z'\sim z_B$ give a contribution to the final asymmetry.

The general expression Eq. (\ref{CPas}) for $\ve_1$ can be
re-cast through the $\O$ matrix as \cite{geometry}
\be\label{ve1}
\ve_1 = \xi(x_2)\,\ve_1^{\rm HL}(m_1,M_1,\O_{21},\O_{31})+
[\xi(x_3)-\xi(x_2)]\,
\D\ve_1 (m_1,M_1,\O_{21},\O_{31},\O_{22}) \, .
\ee
In the HL, for $x_3,x_2\gg 1$, one has
$\xi(x_2)\simeq \xi(x_3) \simeq 1$ and then,
from Eq. (\ref{ve1}), $\ve_1=\ve_1^{\rm HL}(m_1,M_1,\O_{21},\O_{31})$.
Therefore, in the HL the dependence on four of the
see-saw parameters, namely $M_2,M_3,\O_{22}$, cancels out
and one is left only with six parameters. Notice moreover
that $\k_1^{\rm f}=\k_1^{\rm f}(m_1,M_1,K_1)$,
where $K_1=K_1(m_1,\O_{21},\O_{31})$,
and thus the final asymmetry depends on the same six parameters too.

The HL  for $\ve_1$ can be written  as \cite{di02}
$
\ve_1^{\rm HL}(m_1,M_1,\O_{21},\O_{31}) \equiv \,
\overline{\ve}(M_1)\,\b(m_1,\O_{21},\O_{31})
$ ,
where, writing $\O^2_{ij} \equiv X_{ij}-i\,Y_{ij} \equiv
\r_{ij}\,e^{i\,\varphi_{ij}}$, with $\r_{ij} \equiv
|\O^2_{ij}|\geq 0$, one has
\footnote{In \cite{geometry} a minus sign in the
expression for $\b(m_1,\O_{21},\O_{31})$ is missing.
This does not affect any of the results but the quantities
$Y_{i1}$ should be defined as
$Y_{i1} \equiv - {\rm Im}(\O_{j1}^2)$ as here.}
\be
\overline{\ve}(M_1)\,
\equiv {3\over 16\,\pi}\,{M_1\,m_{\rm atm}\over v^2}\,
\hspace{8mm} \mbox{and} \hspace{8mm}
\b(m_1,\O_{21},\O_{31})\equiv
{\sum_j\,m^2_j\,Y_{j1}
\over m_{\rm atm}\,\sum_j\,m_j\,\rho_{j1}} \, .
\ee
It is interesting that $\b(m_1,\O_{21},\O_{31})\leq 1$,
so that in the HL one has the upper bound
$|\ve_1^{\rm HL}|\leq\overline{\ve}(M_1)$ \cite{asaka,di02}.
More precisely one can define an {\em effective phase} $\d_L^{(1)}$ by
\be
\b(m_1,\O_{21},\O_{31})=
\b_{\rm max}(m_1,\mt)\,\sin\d_L^{(1)}(m_1,\O_{21},\O_{31}) \, ,
\ee
such that the upper bound \cite{di02,bound2}~
\be\label{betamax}
\b_{\rm max}(m_1,\mt)= {m_{\rm atm}\over m_1+m_3}\,f(m_1,\mt) \leq 1 \, ,
\ee
corresponds to \mbox{$\sin\d_L^{(1)}=1$} and
it is obtained by maximizing over the $\O$-parameters for fixed $\mt$.
The function $f(m_1,\mt)$ is \cite{geometry}
\be
f(m_1,\mt)={m_1+m_3\over \mt}\,Y_{\rm max}(m_1,\mt) \, ,
\ee
where $Y_{\rm max}(m_1,\mt)$ is the maximum of $Y_{31}$ for $\O_{21}=0$.
For hierarchical light neutrinos, $m_1\lesssim 0.2\,m_{\rm atm}$,
an approximate explicit expression is  \cite{bound2}
\be\label{attempt}
f(m_1,\mt)={m_3-m_1\,\sqrt{1+{m^2_3-m^2_1\over\mt^2}}\over
m_3-m_1} \, ,
\ee
which further simplifies to $f(m_1,\mt)=1-m_1/\mt$
for $m_1 \ll 0.1\,m_{\rm atm}$.
Conversely, in the
quasi-degenerate limit, one has
$f(m_1,\mt)=\sqrt{1-\left({m_1/\mt}\right)^2}$ \cite{hambye}.
One can then conclude that the maximum of the $C\!P$ asymmetry
is reached in the limit $m_1\rightarrow 0$, when $f(m_1,\mt)=1$ and,
since this is true also for $\k_1^{\rm f}$,
it applies also to the final asymmetry $N_{B-L}^{\rm f}\simeq\ve_1\,\k_1^{\rm f}$.
For $\O_{21}=0$ and $Y_{31}=Y_{\rm max}(m_1,\mt)$,
the phase is maximal, while
for a generic choice of $\O$,
the $C\!P$ asymmetry undergoes a phase suppression
\be\label{sind}
\sin\d_L^{(1)}(m_1,\O_{21},\O_{31})={m_1+m_3\over \mt\,f(m_1,\mt)}\,
{(Y_{31}+\sigma^2\,Y_{21})}=
{(Y_{31}+\sigma^2\,Y_{21})\over Y_{\rm max}(m_1,\mt)}\, ,
\ee
where $\s \equiv \sqrt{m_2^2-m_1^2}/m_{\rm atm}$. One can see that
$\sin\d_L^{(1)}=1$ for $Y_{21}=0$ and $Y_{31}=Y_{\rm max}$.
It is instructive to calculate $\sin\d_L^{(1)}$ for each of the three
elementary complex rotations that can be used to parameterize
$\O$ (cf. \ref{second})):
\begin{itemize}
\item For $\O=R_{13}$, one has $\sin\d_L^{(1)}=Y_{31}/Y_{\rm max}$
and the phase is maximal if
\mbox{$Y_{31}=Y_{\rm max}=\mt/m_{\rm atm}$}; notice that
there is no difference between normal and inverted hierarchy.
\item For $\O=R_{12}$, one has
$\sin\d_L^{(1)}=\s^2\,Y_{21}/Y_{\rm max}\leq \s$, larger
for inverted hierarchy compared to the normal one; however
since $\k_1^{\rm f}\propto K_1^{-1.2}$ \cite{proc}
and since $K_1=K_{\rm sol}\,|\O_{21}^2|$ for normal hierarchy
and $K_1\simeq K_{\rm atm}\,|\O_{21}^2|$ for inverted hierarchy,
the final asymmetry is slightly higher for normal
hierarchy compared to the inverted one for fixed $|\O_{21}^2|$~\cite{geometry}.
\item For $\O=R_{23}$, one has $\sin\d_L^{(1)}=\ve_1=0$; one can
check that $\ve_1=0$ applies independently of
$M_2$ and $M_3$ and therefore not only in the HL. Notice that
the conclusions in the previous two cases are still valid if
one multiplies $R_{13}$ or $R_{12}$ with $R_{23}$ respectively,
since it does not affect $\sin\d_L^{(1)}$.
\end{itemize}
Interesting constraints follow
if one imposes that the asymmetry produced from leptogenesis
explains the measured value from
WMAP plus SLOAN combined determination~\cite{cmb,WMAPSLOAN},
\be
\eta_B(m_1,M_1,\O_{21},\O_{31})=
\eta_B^{\rm CMB} = (6.3 \pm 0.3)\times 10^{-10} \, .
\ee
If $M_1 \ll 10^{14}\,{\rm GeV}\,(m_{\rm atm}^2/\sum_i\,m_i^2)$, then
\bea
M_1 & \simeq & {N_{\g}^{\rm rec}\over a_{\rm sph}}
           \,{16\,\pi\,v^2\over 3}\,
          {\eta_B^{\rm CMB}\over m_{\rm atm}}\,
 \left[\k(K_1)\,\b_{\rm max}(m_1,K_1)\,
   \sin\d_L^{(1)}(\O_{21},\O_{31})\right]^{-1} \\
    & \gtrsim & (M_1^{\rm min})_{\rm HL}\equiv  \label{lbM1}
{4.2\times 10^{8}\,{\rm GeV} \over \k(K_1)\,
\b_{\rm max}(m_1,K_1)\,\sin\d_L^{(1)}(\O_{21},\O_{31})}
\hspace{6mm}(\mbox{at}\,\,3\,\s\,\,\mbox{C.L.} ) \; .
\eea
This expression is quite general and shows the effect of the
phase suppression \cite{geometry} and of a higher
absolute neutrino mass scale \cite{proc}
in making the lower bound more restrictive.
In Fig. 4 we show $M_1^{\rm min}$ (thick solid line)
for fully hierarchical light neutrinos ($m_1=0$)
and maximal phase ($\sin\d_L=1$).
\begin{figure}
\centerline{\psfig{file=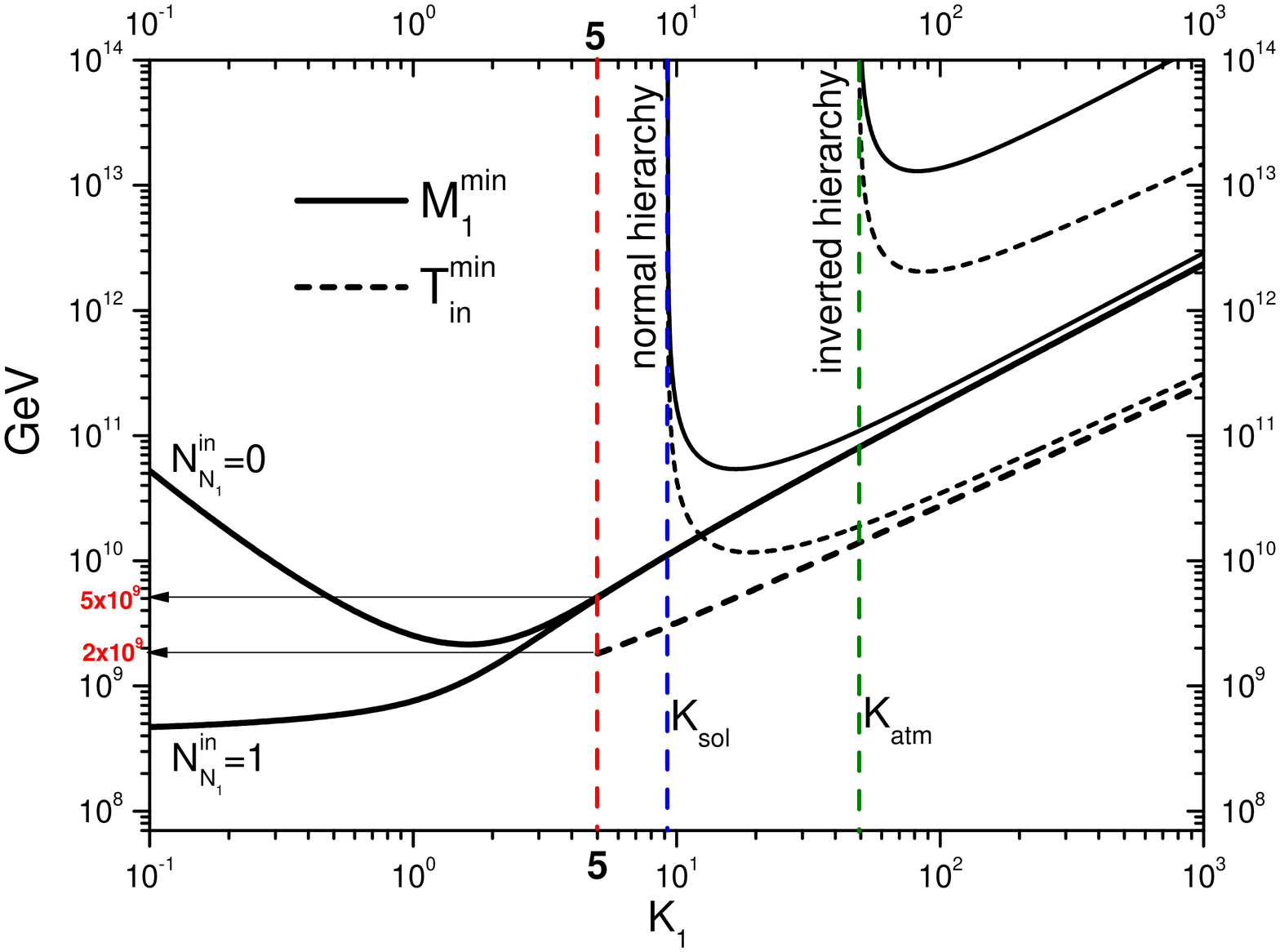,height=80mm,width=13cm}}
\caption{Lower bounds on $M_1$ and $T_{\rm in}$ vs.
$K_1$ (cf. (\ref{lbM1}) and (\ref{lbTin})). Thick lines: case
of maximal phase ($\sin\d_L^{(1)}=1$). Thin lines:
case of very heavy $N_3$.}
\end{figure}
For $K_1\gtrsim 5$ one obtains the lowest value independent
of the initial conditions \cite{geometry},
\be\label{lbM1min}
M_1\gtrsim 5\times 10^{9}\,{\rm GeV} \, .
\ee
The lower bound on $M_1$ also translates into a lower bound on $T_{\rm in}$,
\be\label{lbTin}
T_{\rm in}\geq (T_{\rm in}^{\rm min})_{\rm HL}\simeq
{(M_1^{\rm min})_{\rm HL}\over z_B(K_1)-2}
\gtrsim 2\,\times 10^{9}\,{\rm GeV}
\hspace{8mm} (K_1\gtrsim 5) \, .
\ee
A plot of this lower bound is shown in Fig. 4 (thick dashed line).
The relation between $M_1^{\rm min}$ and $T_{\rm in}^{\rm min}$
can be understood from the top panel of Fig. 3,
showing that the final asymmetry is the result of the decays
occurring just around $z_B$, when inverse decays switch off,
whereas all asymmetry produced before is efficiently washed out.

\section{Beyond the $\mathbf{N_1}$DS}

There are three ways to go beyond the $N_1$DS.
We assume fully hierarchical light neutrinos.

\subsection*{A) $\mathbf{N_2}$DS}

For $\O= R_{23}$, a nice coincidence is realized:
the $C\!P$ asymmetry $\ve_1=0$ while $\ve_2$ can be maximal
and, at the same time, the wash-out from the $N_1$
inverse decays vanishes for $m_1\lesssim 10^{-3}\,{\rm eV}$.
In this way
the final asymmetry can and has to be explained in terms of
$N_2$ decays. A nice feature is that the lower bound on $M_1$
does not hold any more, being replaced by a lower bound on $M_2$
that, however, still implies a lower bound on $T_{\rm in}$ \cite{geometry}.
If one switches on some small $R_{12}$ and $R_{13}$ complex rotations,
then the lower bounds on $M_2$ and $T_{\rm in}$ become
necessarily more stringent.
 Therefore, there is a border beyond which this scenario is
not viable and one is forced to go back to the
usual $N_1$DS for successful leptogenesis.
Account of flavor effects contribute to enlarge the
domain where the $N_2$DS works, since the wash-out from the
lightest RH neutrino is diminished \cite{vives}.
Notice moreover that, like in the $N_1$DS,
the lower bounds are more stringent for inverted
hierarchy than for normal hierarchy, since in the first
case $K_2\gtrsim K_{\rm atm}$, while in the second
case $K_2\geq K_{\rm sol}$.

\vspace{-4mm}
\subsection*{B) Beyond the HL}

If the RH neutrino masses are sufficiently close
    then the $B-L$ asymmetry and the wash-out from the two heavier
    RH neutrinos have also to be taken into account and
     the general expression (\ref{NBmL}) for $N_{B-L}^{\rm f}$
     has to be used; at the same time, the general expression
       Eq. (\ref{ve1}) for $\ve_1$ has also to be used.
       Here the first, typically dominant, term can be
      enhanced by a factor $\xi(x_2)$ \cite{bound2} while
  the second term can be calculated using \cite{geometry}
   \be
    \Delta\ve_1\equiv {3\over 16\,\pi}\,{{\rm Im}
     [(h^{\dagger}\,h)^2_{13}]\over
     (h^{\dagger}h)_{11}}
      \,{1\over \sqrt{x_3}} = \overline{\ve}(M_1)\,
       {{\rm Im}\left[\sum_h\,m_h\,\O_{h1}^{\star}\O_{h3}\right]^2\over
         m_{\rm atm}\,\mt}  \, .
    \ee
    One can see that $\D\ve_1$ vanishes for $\O=R_{12}$. It also vanishes
      in another interesting situation, the
        {\em strong-strong wash-out scenario} for very large $M_3$,
       that will be discussed in detail in the next Section.
          In these two interesting cases, the
         enhancement $\xi(x_2)$ is the only effect on $\ve_1$ and
         the HL is recovered, within a 10\% precision, for
           $\d_2 \gtrsim 5$ \cite{bound2,geometry}.
       These two cases have also been recently studied in
        \cite{branco} for $\d_2\ll 1$, within the context
         of radiative leptogenesis.

It is interesting that even the enhancement
$\xi(x_2)$ can be absent in a particular case.
This happens if one considers fully hierarchical light neutrinos and
$\O=R_{13}$. It is easy to calculate that in this case one has
$\D\ve_1=\overline{\ve}(M_1)\,\sin\d_L^{(1)}$.
Plugging this term into  the general expression (\ref{ve1}) for $\ve_1$,
one has an exact cancellation of the terms proportional to $\xi(x_2)$
and in the end
\be
\ve_1|_{\O=R_{13}} = \xi(x_3)\,\overline{\ve}(M_1)\, \sin\d_L^{(1)} \, .
\ee
For $x_3\gg 1$, such that $\xi(x_3) \simeq 1$,
there is no enhancement of the usual $C\!P$ asymmetry when
$M_2\rightarrow M_1$.  This example disproves the misconception
that degenerate RH neutrino masses unavoidably lead to $C\!P$
asymmetry enhancement. Moreover, it shows that the usual
most stringent lower bounds on $M_1$ and $T_{\rm in}$, for
\mbox{$\sin\d_L^{(1)}=1$}, continue to be valid even beyond the HL.
However, they can be evaded in other models, which means
that, in general, changing the RH neutrino mass spectrum,
the $\O$ matrix that maximizes the asymmetry changes too.

As a last exercise, one can check that $\D\ve_1$, like
$\ve_1^{\rm HL}$, vanishes for $\O=R_{23}$, confirming that
the $N_2$DS is the only possibility for this particular choice.

\subsection*{C) Large $\mathbf{|\O_{22}|}$}

There is a situation
when the term $[\xi(x_3)-\xi(x_2)]\,\D\ve_1$ becomes dominant
over $\xi(x_2)\,\ve_1^{\rm HL}$ and one has to require
a stronger hierarchy to recover the HL \cite{hambye}. This
term is maximized over $x_3$ for $x_3\gg 1$, implying
$\xi(x_3) \simeq 1$.
Moreover, if for definiteness one
imposes $\O_{21}=X_{31}=0$,
so that $\sin\d_L^{(1)}=1$, then \cite{geometry}
\be
\xi_{\ve_1}\equiv {\ve_1\over\overline{\ve}(M_1)}
= \xi(x_2)+[1-\xi(x_2)]\,(X_{22}+Y_{\rm max}\,Y_{22}) \, ,
\ee
where $Y_{\rm max}=\mt/m_{\rm atm}$.
One can see that if $X_{22}=1$ and $Y_{22}=0$,
corresponding to $\O=R_{13}$, then one recovers $\xi_{\ve_1}=1$,
independently of the value of $x_2$.
However, one can now perceive another possibility:
if $|X_{22}+Y_{\rm max}\,Y_{22}|\gg 1$, then the $C\!P$
asymmetry can be enhanced, i.e. $\xi_{\ve_1}>1$,
even in the HL, when $x_2\gg 1$.
There are nevertheless some limitations. First of all,
Yukawa couplings cannot be larger than $\sim 0.1$
for the Eq. (\ref{CPas}) to hold.
Moreover, when calculating the final $B-L$ asymmetry,
one also has to take into account the increased wash-out.
In the end, for fully hierarchical neutrinos, this possibility
should be regarded as a very special case, also because
it implies unnaturally huge phase cancellations
due to the $\O$ orthogonality. On the other hand,
for quasi-degenerate light neutrinos
this term can now be more easily dominant, since the
first term is suppressed by $\b_{\rm max}(m_1,\mt)<1$ \cite{di02,bound2},
while $\D\ve_1$ is not and thus an account of this term
makes possible to evade the upper bound on the light
neutrino masses \cite{hambye}.


\section{Beyond the HL in strong-strong wash-out scenarios}

In this section we discuss the effects arising in models beyond the HL
and the conditions for the HL to be recovered.
There is a large variety of possibilities and for definiteness we focus
on a particularly  interesting class that
provides a useful framework to understand the general effects.
We still assume a partial hierarchy, such that $M_3\gg M_1,M_2$,
while $M_1$ and $M_2$ can be arbitrarily close.
This results in $|\ve_3|\ll |\ve_1|,|\ve_2|$ and the final asymmetry
can be calculated as
\be\label{NBmLfss}
N_{B-L}^{\rm f}\simeq \ve_1\,\k_1^{\rm f}+\ve_2\,\k_2^{\rm f} \, .
\ee
Another convenient restriction is to focus on
{\em strong-strong (ss) wash-out scenarios}, where both
$K_1$ and $K_2 \gtrsim 5$. In this way
$\k_1^{\rm f}$ and $\k_2^{\rm f}$ can
be calculated inserting the equilibrium values for both
rate abundances into the general expressions (\ref{ki}), such that
\bea \label{k1ss}
\k_1^{\rm f}(K_1,K_2,\d_2)  \simeq   \k_1^{\rm ss}(K_1,K_2,\d_2)
& \equiv & - \int_{0}^{\infty}\, dz'\
{dN_{N_1}^{\rm eq}\over dz'}\,
e^{-\,\int_{z'}^{\infty}\ dz''\, [W_{1}^{\rm ID}(z'')+W_2^{\rm ID}(z'')]}\, ,
\\  \label{k2ss}
\k_2^{\rm f}(K_1,K_2,\d_2)  \simeq   \k_2^{\rm ss}(K_1,K_2,\d_2)
& \equiv & - \int_{0}^{\infty}\, dz'\,
{dN_{N_2}^{\rm eq}\over dz'}\
e^{-\,\int_{z'}^{\infty}\ dz''\, [W_{1}^{\rm ID}(z'')+W_2^{\rm ID}(z'')]} \, .
\eea
In Fig. 5 we have plotted $\k_1^{\rm ss}(K_1,K_2,\d_2)$
and $\k_2^{\rm ss}(K_1,K_2,\d_2)$ for the indicated values of $\d_2$,
and for $K_2=K_{\rm atm}$. This value  is particularly convenient
to highlight the cumulative effect of the wash-out.
\begin{figure}
\centerline{\psfig{file=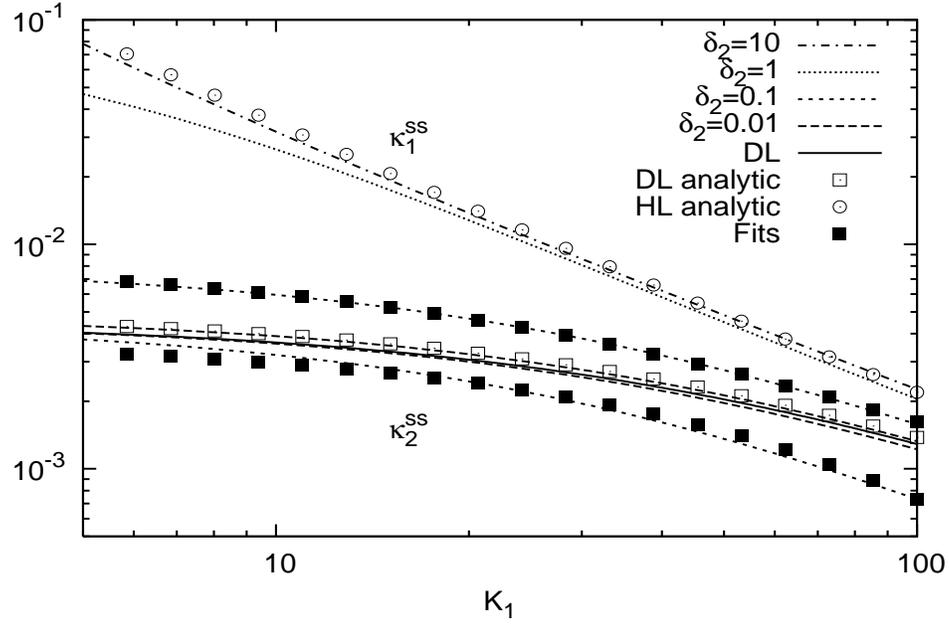,height=9cm,width=14cm}}
\centerline{\psfig{file=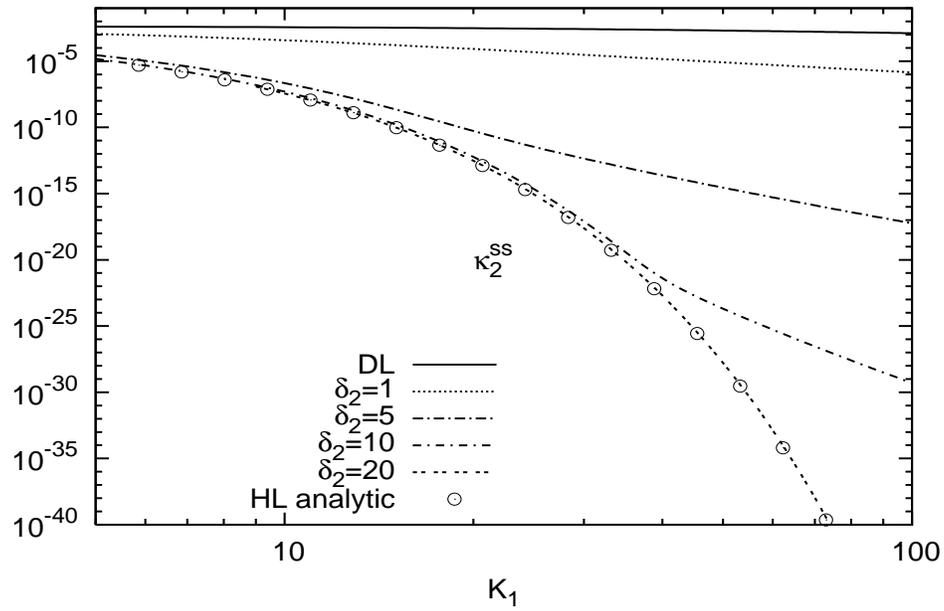,height=9cm,width=14cm}}
\caption{Efficiency factors in the strong-strong wash-out case
for $K_2=K_{\rm atm}$.}
\end{figure}

There are two simple limits where these integrals can be worked out explicitly.
The first is the HL ($\d_{2}\gg 1$). In this case the wash-out from
$N_2$ inverse decays acts only at $z\ll z_B-2$ and does not affect
the dominant contribution produced around $z\sim z_B$.
Therefore, one recovers the result valid in the
$N_1$DS, i.e. $\k_1^{ss}|_{\rm HL}\simeq \k(K_1)$ (cf.~(\ref{k})).
It is also possible to obtain an analytic expression
for $\k_2^{\rm ss}$ in the HL considering that
the wash-out from $N_1$ inverse decays does not interfere with the
asymmetry production and with the wash-out from $N_2$ decays
and inverse decays respectively.
Indeed, the wash-out from $N_1$ inverse decays is active for
$z_{\rm on}(K_1)\simeq 2/\sqrt{K_1}\lesssim z \lesssim z_{\rm off}(K_1)$
and peaked at $z_{\rm max}\simeq 2.4$ \cite{annals}, when
the wash-out by $N_2$ inverse decays is already off at
$z\simeq z_{\rm off}(K_2) M_1/M_2 \ll z_{\rm on}(K_1)$ and
practically all $N_2$'s have already decayed.
This results in a simple factorized analytic expression,
\be\label{HL}
\k_2^{\rm ss}|_{\rm HL}\simeq \k(K_2)\,
e^{-\,\int_{0}^{\infty}\ dz'\,W_{1}^{\rm ID}(z')}
\simeq \k(K_2)\,e^{-{3\,\pi\,\over 8}\,K_1} \, ,
\ee
that is shown in Fig. 5 together with Eq. (\ref{k}) for $\k_1$
(circles). One can see how they well reproduce $\k_1^{\rm ss}$ and
$\k_2^{\rm ss}$ in the HL.
For $\d_2>\d_2^{\rm HL}|_{\k_1^{\rm f}}$
the HL for $\k_1^{\rm f}$ is recovered within 10\%.
In Fig. 6 we show $\d_2^{\rm HL}|_{\k_1^{\rm f}}$ as a function of $K_1$
for fixed $K_2=K_{\rm atm}$. We also
show (crosses) an analytic conservative estimate \cite{geometry},
\be\label{d2HL}
\d_2\geq  {z_B(K_2)+2\over z_B(K_1)-2}-1 \, ,
\ee
obtained neglecting the wash-out of the lightest
RH neutrino on the asymmetry produced from the second lightest. One can see
how indeed this is always more conservative, especially at large values of
$K_1$, where the wash-out of the lightest RH is stronger.
\begin{figure}
\centerline{\psfig{file=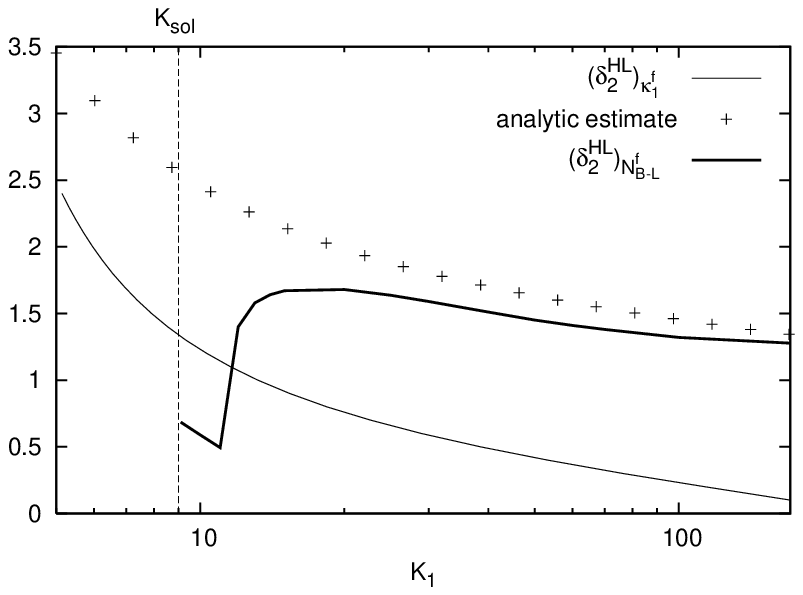,height=9cm,width=14cm}}
\caption{Plot of $\d_2^{\rm HL}|_{\k_1^{\rm f}}$
and of $\d_2^{\rm HL}|_{N_{B-L}^{\rm f}}$ vs. $K_1$
for $K_2=K_{\rm atm}$.
The crosses are the analytic estimate Eq. (\ref{d2HL}).}
\end{figure}
The second limit is the DL ($M_2=M_1$, i.e. $\d_2=0$).
This has been recently studied
in \cite{branco} within the context of radiative
leptogenesis. In this case one has $z_2=z_1=z$, so that
$dN_{N_1}^{\rm eq}/dz'=dN_{N_2}^{\rm eq}/dz'$
and $W_1^{\rm ID}/K_1 = W_2^{\rm ID}/K_2$ (cf. (\ref{WID})).
It is then easy to derive a simple result,
\be\label{DL}
\k_1^{ss}(K_1,K_2)|_{\d_2=0}=
\k_2^{ss}(K_1,K_2)|_{\d_2=0}=
\k(K_1+K_2) \, ,
\ee
indicating that in the DL the $N_1$ and $N_2$ wash-out contributions
add up and tend to suppress the final asymmetry compared to the HL.
This effect goes into the opposite direction compared to the enhancement
of the $C\!P$ asymmetry when masses get close. Therefore one has to
be careful and check which  effect is dominant between the two.
As we will see a precise answer depends on the particular
form of the orthogonal matrix and on the values of $\d_2$,
$K_1$ and $K_2$. One can see in Fig.~5 that the expression
(\ref{DL}) reproduces well both $\k_1^{\rm ss}$ and $\k_2^{\rm ss}$
in the DL (open squares). Notice that this result is easily extended
to the full DL ($M_1=M_2=M_3$) with all $K_i\gtrsim 5$,
where one obtains $\k_i^{\rm f}=\k(K_1+K_2+K_3)$, $i=1,2,3$.

We have also worked out useful fits for any value of $\d_2$,
\be
\k_1^{\rm fit}(\d_2)={2\over z_B(K_1+K_2^{(1-\d_2)^3})\,
(K_1+K_2^{1-\d_2})}
\ee
and
\be
\k_2^{\rm fit}(\d_2)=
2\,
{\left[1+2\,\ln\left({1+\d_2\over 1-\d_2}\right)\right]^2
\over
z_B(K_2+K_1^{(1-\d_2)^3})\,
(K_2+K_1^{1-\d_2})
}\,
\times e^{-{3\,\pi\over 8}\,K_1\,
\left({\d_2\over 1+\d_2}\right)^{2.1}} .
\ee
In Fig.~5 one can see (black squares) that they fit well
$\k_1^{\rm ss}$ and $\k_2^{\rm ss}$ for $\d_2=0.1$.
Notice moreover that the analytic expressions (\ref{k}), (\ref{HL})
are correctly recovered in the HL and (\ref{DL}) in the DL.

We have thus seen that the simple class of strong-strong wash-out scenarios
allows one to describe the most relevant features of models beyond the HL
with simple analytic expressions. More general cases can be
easily understood and studied extending these results.
In order to calculate the $C\!P$ asymmetries and the
lower bound on $M_1$ in scenarios beyond the HL, we focus
on two particularly interesting realizations
of strong-strong wash-out scenarios: the first is $\O=R_{13}$,
where the phase is maximal and an interesting result holds for the $C\!P$
asymmetry $\ve_1$ beyond the HL,
the second is the limit $M_3\gg 10^{14}\,{\rm GeV}$,
where the heaviest RH neutrino decouples
and plays no role.

\subsection{`Maximal phase' scenario}

Assuming $\O=R_{13}$ (cf. (\ref{R})) and $M_3\gg M_1,M_2$
we have seen that the $C\!P$ asymmetry $\ve_1$, for
hierarchical light neutrinos, is always given by the HL
independently of $\d_2$, i.e. $\ve_1=\bar{\ve}(M_1)\,\sin\d_L^{(1)}$,
where the phase is maximal if $X_{31}=0$.
Moreover $\ve_2=0$, so that the asymmetry
is generated only from $N_1$ decays. The only effect
of $N_2$ is the additional wash-out, described by
a fixed value of the decay parameter $\bar{K}_2=K_{\rm sol}$
for normal hierarchy and
$\bar{K}_2=\sqrt{K_{\rm atm}^2-K_{\rm sol}^2}$
for inverted hierarchy.
The final asymmetry  is then, for maximal phase, given by
\be
N_{B-L}^{\rm f}(M_1,K_1,\d_2) =
\bar{\ve}(M_1)\,
\k_1^{\rm f}(K_1,\bar{K}_2,\d_2) \, .
\ee
For $K_1\gtrsim 5$ one can use for $\k_1^{\rm f}$
the analytic expressions obtained in the strong-strong
wash-out regime and one obtains for the lower bounds,
\be
M_1^{\rm  min}(K_1,\d_2)\simeq
{4.2\times 10^{8}\,{\rm GeV}\over \k_1^{\rm ss}(K_1,\bar{K}_2,\d_2)} \, ,
\ee
and
\be
T_{\rm in}^{\rm min}(K_1,\d_2)=
{M_1^{\rm  min}(K_1,\d_2)\over z_{\star}(K_1,\bar{K}_2,\d_2)}\, ,
\ee
where $z_{\star}(K_1,K_2,\d_2)$ is
the $T_{\rm in}^{\rm min}$ relaxation factor and is
approximately given, in the HL, by
$z_{\star}^{\rm HL}\simeq z_B(K_1)-2$ \cite{annals} and, in the DL,
by $z_{\star}^{\rm DL}\simeq z_B(K_1+K_2)-2$.
For intermediate situations a good fit is given by
$z_{\star}(K_1,K_2,\d_2)\simeq z_B(K_1+K_2^{(1-\d_2)^3})-2$.
In the top panel of Fig. 8 we compare the lower bounds
valid in the HL with those for $\d_2=0.1$ (thick lines).
One can see how the
effect of the additional wash-out makes them more stringent.
Moreover, one can see how the degeneracy between normal
and inverted hierarchy in the HL is broken for finite value
of $\d_2$ because of the different $K_2$ value. In the
bottom panel we have compared the lower bounds for $\d_2=0.1$
with those for $\d_2=0.01$ (thick lines). One can see that
these sensibly change only at small $K_1\sim 10$, while at
larger values they are the same. This results from the fact
that the effect of additional wash-out saturates
and the DL is reached (cf. Fig.~5).
Therefore, this model represents an interesting example
of how going beyond the HL does not necessarily relaxes
the lower bounds on $M_1$ and $T_{\rm in}$.

\subsection{The limit of very heavy $\mathbf{N_3}$}

In the limit $M_3\gg 10^{14}\,{\rm GeV}$, the orthogonal see-saw matrix
$\O$ necessarily reduces to a special form \cite{fgy,ir,ct}
\be\label{ss}
\O=
\left(
\begin{array}{ccc}
     0            &   0                 &  1 \\
\sqrt{1-\O^2_{31}}& -\O_{31}            &  0 \\
       \O_{31}    & \sqrt{1-\O^2_{31}}  &  0
\end{array}
\right)
\, ,
\ee
obtained from the general one for
$\O^2_{22}=\O^2_{31}$ and $\O^2_{22}+\O^2_{31}=1$. In terms of
the parametrization with complex rotations this corresponds
to $w_{22}=0$ and $w_{21}=1$, implying $\O_{21}=\sqrt{1-w^2_{31}}$
(cf. (\ref{rel})).
This limit also implies \mbox{$m_1\sim 0.01\,{\rm eV}\,
(10^{14}\,{\rm GeV}/M_3)\,
({\rm Re}[U^{\dagger}\,h]^2_{13}/0.1)\ll m_{\rm sol}$}, i.e. fully hierarchical
light neutrinos.
For normal hierarchy the values of $K_1$ and $K_2$ are given by
\be
K_1 =  K_{\rm sol}\,\r_{21}+ K_{\rm atm}\,\r_{31} \,
\hspace{8mm}\mbox{and}\hspace{8mm}
K_2  = K_{\rm sol}\,\r_{31}+ K_{\rm atm}\,\r_{21} \, ,
\ee
so that $K_1,K_2\geq K_{\rm sol}\simeq 9$ and
$K_1+K_2 \geq K_{\rm sol}+K_{\rm atm}\simeq 60$.
For inverted hierarchy the same expressions hold with the replacement
$K_{\rm sol}\rightarrow \sqrt{K_{\rm atm}^2-K_{\rm sol}^2}$.
Therefore, in this scenario, both $N_1$ and $N_2$
decay in the strong wash-out regime, while the heaviest RH
neutrino decouples completely.
Notice moreover that  $\ve_3=0$, so that
the final $B-L$ asymmetry can be calculated as the sum of
the two contributions from the two lightest RH neutrinos
and it will depend on a set of 4 parameters that can be
conveniently chosen to be $M_1,\, K_1 ,\, K_2$ and $\d_2$.

Let us calculate the two $C\!P$ asymmetries $\ve_1$ and $\ve_2$.
As anticipated, it is easy to check that $\Delta\ve_1$,
defined in Eq.~(\ref{ve1}) for $\ve_1$, vanishes.
Therefore, the $C\!P$ asymmetry enhancement is described
just by the function $\xi(x_2)$, as in \cite{bound2}.
Moreover, since light neutrinos are fully hierarchical,
one has $\b_{\rm max}(m_1,\mt)=1$ (cf. (\ref{betamax})).
The expression (\ref{sind}) for the effective phase
$\sin\d_L^{(1)}$ specializes to
\be\label{sind1}
\sin\d_L^{(1)}=
{K_{\rm atm}\over K_1}\,Y_{31}\,(1-\s^2) \, ,
\ee
where $Y_{31}$ has to be regarded as a function of $K_1$ and $K_2$.
Turning to $\ve_2$, the general expression (\ref{CPas}) can be re-cast as
\be
\ve_2={3\over 16\,\pi}\,\sum_{i=1,3}\,
{{\rm Im}[(h^{\dagger}\,h)^2_{2i}]\over (h^{\dagger}h)_{22}}
{\xi(x_i/x_2)\over \sqrt{x_i/x_2}} \, ,
\ee
and using \cite{geometry}
\be
(h^{\dagger}\,h)_{ij}={\sqrt{M_i\,M_j}\over v^2}\sum_h\,\,m_h\,
\O_{hi}^{\star}\,\O_{hj} \, ,
\ee
one can check that the term $(h^{\dagger}\,h)_{23}=0$.
After some algebraic manipulations, one finds
\be
\ve_2\simeq \bar{\ve}(M_2)\,\xi\left({1/x_2}\right)\,\sin\d_L^{(2)} \, .
\ee
We have introduced a second effective leptogenesis phase
\be
\sin\d_L^{(2)}\equiv \, -{K_{\rm atm}\over K_2}\,Y_{31}\,(1-\s^2) =
-{K_1\over K_2}\,\sin\d_L^{(1)}  \, ,
\ee
which always has opposite sign compared to $\sin\d_L^{(1)}$.
The final asymmetry  (cf. (\ref{NBmLfss})) can then be written as
\be\label{NBmLfss2}
N_{B-L}^{\rm f}(M_1,K_1,K_2,\d_2)  =
\bar{\ve}(M_1)\,\sin\d_L^{(1)}(K_1,K_2)\,\a(K_1,K_2,\d_2) \, ,
\ee
where
\be\label{alpha}
\a(K_1,K_2,\d_2)\equiv
\xi(x_2)\,\,\k_1^{\rm ss}(K_1,K_2,\d_2)-\sqrt{x_2}\,\,
\xi\left({1/x_2}\right)\,{K_1\over K_2}\,\k_2^{\rm ss}(K_1,K_2,\d_2) \, .
\ee
Notice that in the HL one has
\be
{\ve_2\,\k_2^{\rm ss} \over \ve_1^{\rm HL}\,\k_1^{\rm ss}|_{\rm HL}}  \simeq
-\sqrt{x_2}\,\,\xi\left({1/ x_2}\right)\,{K_1\over K_2}\,
{\k_2^{\rm ss}(K_1,K_2,\d_2) \over \k(K_1)}
  \stackrel{\rm HL}{\longrightarrow}
-{2\,\ln x_2 \over 3\,\sqrt{x_2}}\,\,
{K_1^2\over K_2^2}\,{z_B(K_1)\over z_B(K_2)}
e^{-{3\,\pi\over 8}\,K_1}  \ll 1 \, .
\ee
As a consequence the contribution from the second lightest RH neutrino
becomes negligible and one recovers the $N_1$DS.
In Fig.~6 we show $\d_2^{\rm HL}|_{N_{B-L}^{\rm f}}$,
such that for $\d_2>\d_2^{\rm HL}|_{N_{B-L}^{\rm f}}$
the HL for the final asymmetry, in the case of very heavy $N_3$,
is recovered within a $10\%$ precision.

In Fig. 7 we give an example of the dynamics
of the asymmetry generation with two RH neutrinos,
a generalization of the example of Fig. 3 for the $N_1$DS.
We show the quantity
$[d\k_1/dz'+(\ve_2/\ve_1)\,d\k_2/dz']_{z'\leq z}$
for $K_1=10$ and $K_2=50$. In the top panel
$\sqrt{x_2}=M_2/M_1=10\,(\d_2=9)$.
For this choice  of the parameters, one has
$\ve_2/\ve_1\simeq -0.035$.
At small values $z\simeq 0.1$ the $N_2$ decays give a non-negligible
contribution to the total asymmetry but with negative sign.
However, this contribution is completely washed-out
at $z\sim 1$ by the $N_1$ inverse decays.
Therefore, this example illustrates well
how the HL is recovered for large $\d_2$, where the production
of the asymmetry and the wash-out from $N_2$ and from $N_1$
occur at two well-separated stages. This numerical example completes the
qualitative discussion in \cite{geometry} where the
condition (\ref{d2HL}) was found.
In the bottom panel we show an example for $\d_2=0.1$, implying
a minus sign in $\xi(1/x_2)$ (cf. Fig. 1)
that cancels with the minus sign in $\sin\d_L^{(2)}$, such that
now the asymmetry produced from the $N_2$'s has the same sign
as the one produced from the $N_1$'s.
Therefore, both the two productions and
the two wash-out occur simultaneously and add up.

\begin{figure}[p]

\vspace{2cm}

\begin{center}
\begin{picture}(100,100)
\put(-130,0){\includegraphics[scale=0.9]{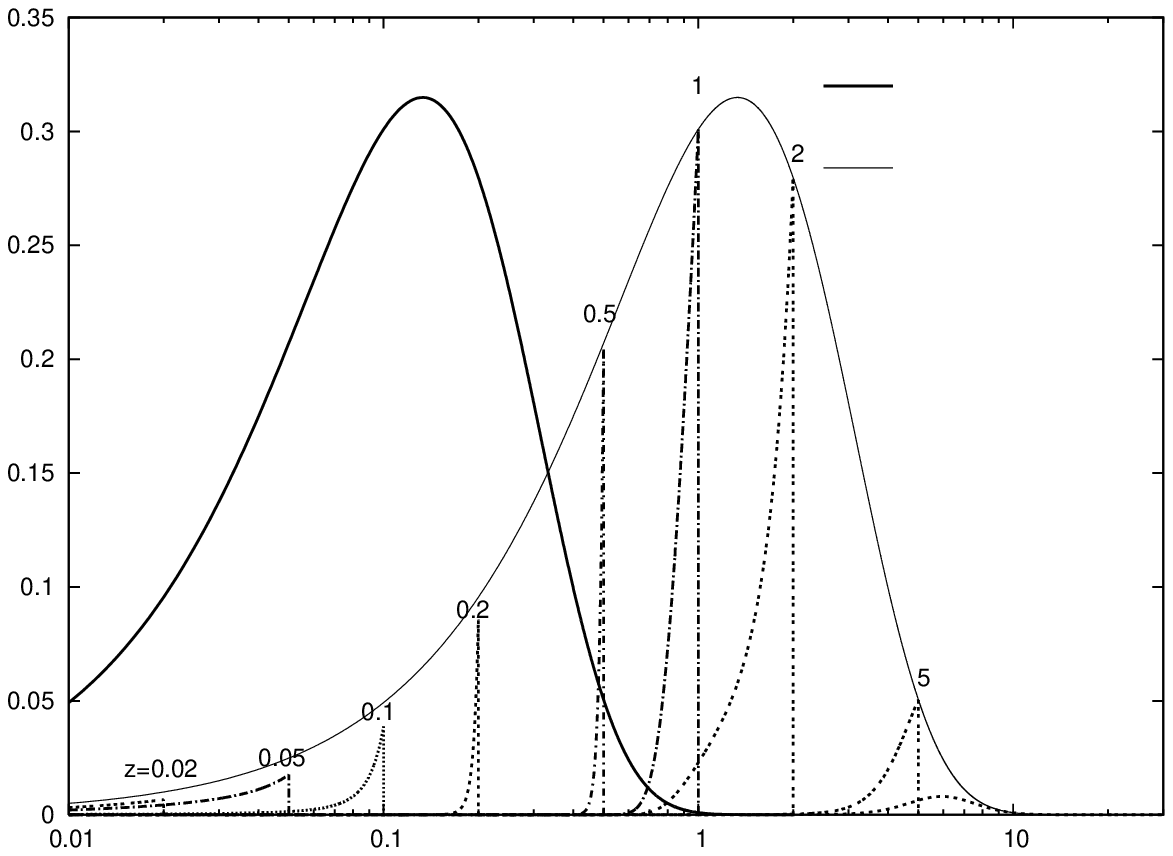}}
\put(-90,200){$K_1=10$}
\put(-90,180){$K_2=50$}
\put(-90,160){$\delta_2=9$}
\put(50,-10){$z'$}
\put(118,200){\tiny $\left| dN_{N_2} \over dz' \right| \simeq
\left|  dN_{N_2}^{eq} \over dz' \right| $}
\put(118,180){\tiny $\left| dN_{N_1} \over dz'\right| \simeq
\left| dN_{N_1}^{eq} \over dz' \right| $}
\put(127,5){$z_B$}
\put(135,32){$z\gg z_B$}
\put(130,19){$\swarrow$}
\put(-150,80)
{
\begin{sideways}
$\left(\frac{dk_1}{dz'} +
\frac{\epsilon_2}{\epsilon_1}\frac{d\k_2}{dz'}\right)_{z' \leq z}$
\end{sideways}
}
\end{picture}

\vspace{6cm}

\begin{picture}(100,100)
\put(-130,0){\includegraphics[scale=0.9]{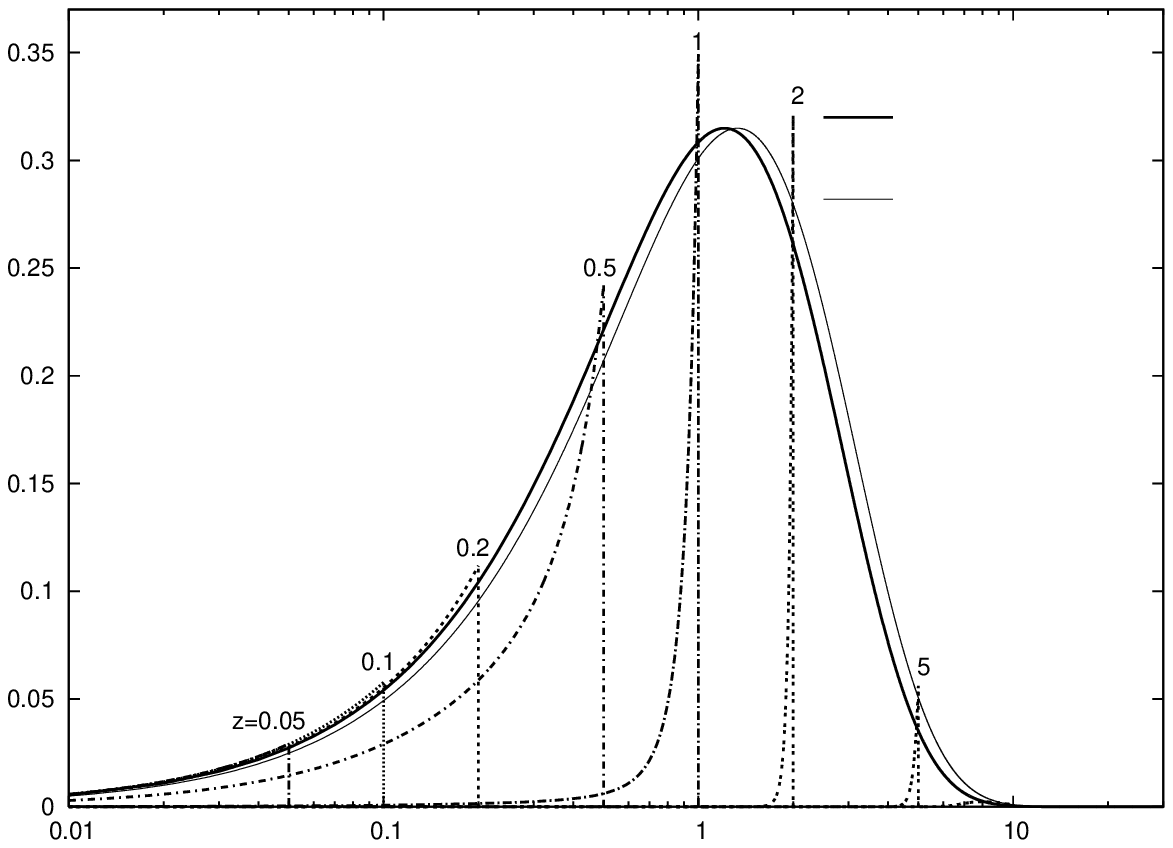}}
\put(-90,200){$K_1=10$}
\put(-90,180){$K_2=50$}
\put(-90,160){$\delta_2=0.1$}
\put(50,-10){$z'$}
\put(118,190){\tiny $\left| dN_{N_2} \over dz' \right| \simeq
\left| dN^{eq}_{N_2} \over dz' \right|$}
\put(118,170){\tiny $\left| dN_{N_1} \over dz'\right| \simeq
\left| dN^{eq}_{N_1} \over dz' \right|$}
\put(128,5){$z_B$}
\put(145,27){$z\gg z_B$}
\put(140,17){$\swarrow$}
\put(-150,80)
{
\begin{sideways}
$\left( \frac{d\k_1}{dz'} +
\frac{\epsilon_2}{\epsilon_1}\frac{d\k_2}{dz'}\right) _{z' \leq z}$
\end{sideways}
}
\end{picture}
\end{center}
\caption{Dynamics of the asymmetry generation with 2 RH neutrinos.}
\end{figure}
Let us now calculate the lower bounds on $M_1$ and on $T_{\rm in}$.
They can be written as
\be
M_1^{\rm  min}(K_1,\d_2)=
{4.2\times 10^{8}\,{\rm GeV}\over
\sin\d_L^{(1)}(K_1,K_2^{\star})\,\a(K_1,K_2^{\star},\d_2)}
\ee
and
\be
T_{\rm in}^{\rm min}(K_1,\d_2)=
{M_1^{\rm  min}(K_1,\d_2)\over z_{\star}(K_1,K_2^{\star},\d_2)}\, ,
\ee
where $K_2^{\star}$ is the value of $K_2$ that
maximizes $N_{B-L}^{\rm f}$ (cf.~(\ref{NBmLfss2})).
In the HL one has
$\a(K_1,K_2^{\star},\d_2\gg 1)=\k_1^{\rm ss}|_{\rm HL}\simeq \k(K_1)$,
such that the $N_1$DS is recovered but the
effective phase is not maximal (cf. (\ref{sind1}))
and the lower bounds are more stringent \cite{ct}.
Maximizing the effective phase $\sin\d_L^{(1)}(K_1,K_2)$
over $K_2$, one obtains
$\sin\d_L^{(1)}=f(m_2,\mt)\,(1-\s)$ \cite{geometry}.
In the case of normal hierarchy
$\s\simeq m_{\rm sol}/m_{\rm atm}\simeq 0.19$.
The function $f(m_{\rm sol},\mt)$ can then be calculated
using Eq. (\ref{attempt}) with the replacement
$m_1\rightarrow m_{\rm sol}\ll m_{\rm atm}$.
In this way, from Eq. (\ref{lbM1}) for $(M_1^{\rm min})_{\rm HL}$,
one obtains
\be
(M_1^{\rm min})_{\rm HL} \simeq {4.2\,\times\, 10^{8}\,{\rm GeV}
\over \k(K_1)\,f(m_{\rm sol},\mt)\,(1-\s)} \, .
\ee
This lower bound and the associated one on $T_{\rm in}^{\rm min}$
are shown in Fig.~4 (thin lines) and compared with
those obtained for $\sin\d^{(1)}_L=1$ (thick lines).
For inverted hierarchy one has
$\s\simeq \sqrt{1-(m_{\rm sol}/m_{\rm atm})^2}$ and
$f(m_2,\mt)\simeq \sqrt{1-(m_2/\mt)^2}$. The
bounds become much more restrictive \cite{ct} for two reasons:
first now $K_1\gtrsim K_{\rm atm}$, second
there is strong phase cancellation in $\sin\d_L^{(1)}$.
This actually  occurs on more general grounds \cite{geometry} and one can
say that for inverted hierarchy the constraints are more stringent
than for normal hierarchy, except for the two special cases
already mentioned in Section 3.
The first one is
for $\O=R_{13}\,R_{23}\, (\O_{21}=0)$ and $Y_{31}=Y_{\rm max}$, where
the phase is maximal and one obtains the most conservative lower bounds on
$M_1$ and $T_{\rm in}$ in the HL, the same for normal and inverted hierarchy.
As soon as one diverges from this special case,
the degeneracy gets broken.
A second special case is for $\O=R_{12}\,R_{23}\,(\O_{31}=0)$
and $Y_{21}=Y_{\rm max}/\s$, so that $\sin\d_L^{(1)}=\s$,
larger for inverted  hierarchy than for the normal one. However,
as we said, the wash-out is stronger for inverted hierarchy than for
normal hierarchy and the two things compensate, such that the lower
bounds are practically equivalent, though not equal. Except for these
two cases the allowed region in the space parameters
is larger for normal hierarchy than for inverted hierarchy
and leptogenesis tends to favors normal over inverted
one. The case of very large $M_3$ represents the most
extreme one in this respect. Also in the $N_2$DS the
lower bounds on $M_2$ and $T_{\rm in}$
are more stringent in the case of inverted
hierarchy than in normal hierarchy. We can thus conclude
that if the light neutrino hierarchy were to be inverted, then
the allowed region would be considerably reduced.

Let us now see what happens beyond the HL, i.e. for small values of $\d_2$.
There are three effects, all accounted for by the function $\a(K_1,K_2,\d_2)$
(cf. (\ref{alpha})). The first is that the asymmetry produced
from the $N_2$'s becomes non-negligible and since $\xi(1/x_2)$
becomes negative at small $\d_2$, cancelling the minus sign in
$\sin\d_L^{(2)}$,  the two asymmetries add up. However,
this effect brings a factor 2 enhancement at most \cite{bound2},
while the major effect is the enhancement of both $\ve_1$ and $\ve_2$,
described by  $\xi(x_2)$ for $\ve_1$ and by $\xi(1/x_2)$ for $\ve_2$.
The third effect is that also the two wash-out effects add up.
As we have seen (cf.~Fig.~5) this effect saturates at $\d_2\simeq 0.01$
and goes into the opposite direction compared to the other two, thus
contrasting the enhancement of the asymmetry for moderately
small values $\d_2\sim 0.1$. This is clearly visible in
the top panel of Fig. 8, where we have compared the results
holding in the HL with those obtained for $\d_2=0.1$.
\begin{figure}
\centerline{\psfig{file=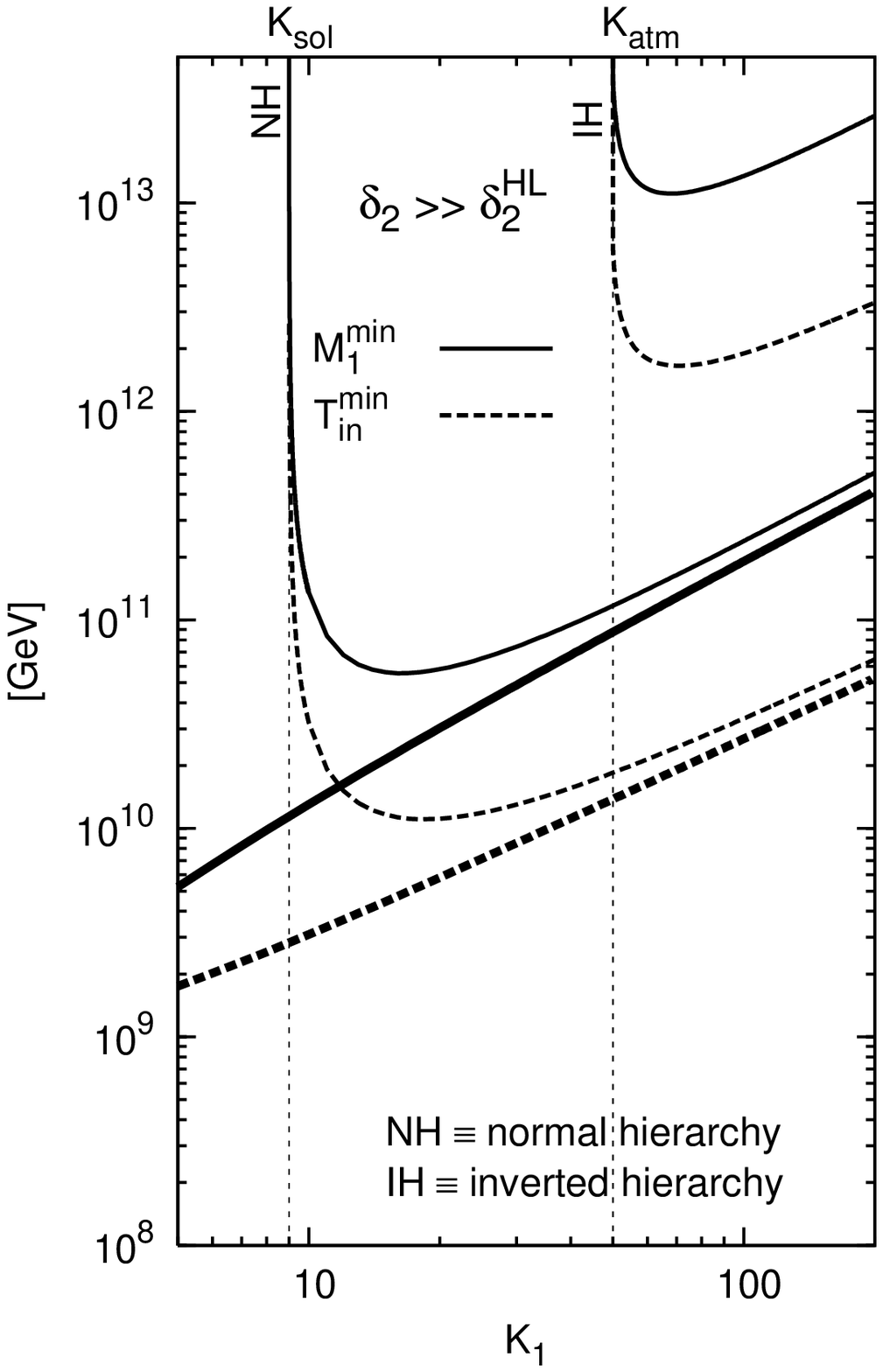,height=85mm,width=75mm} \hspace{-5mm}
\psfig{file=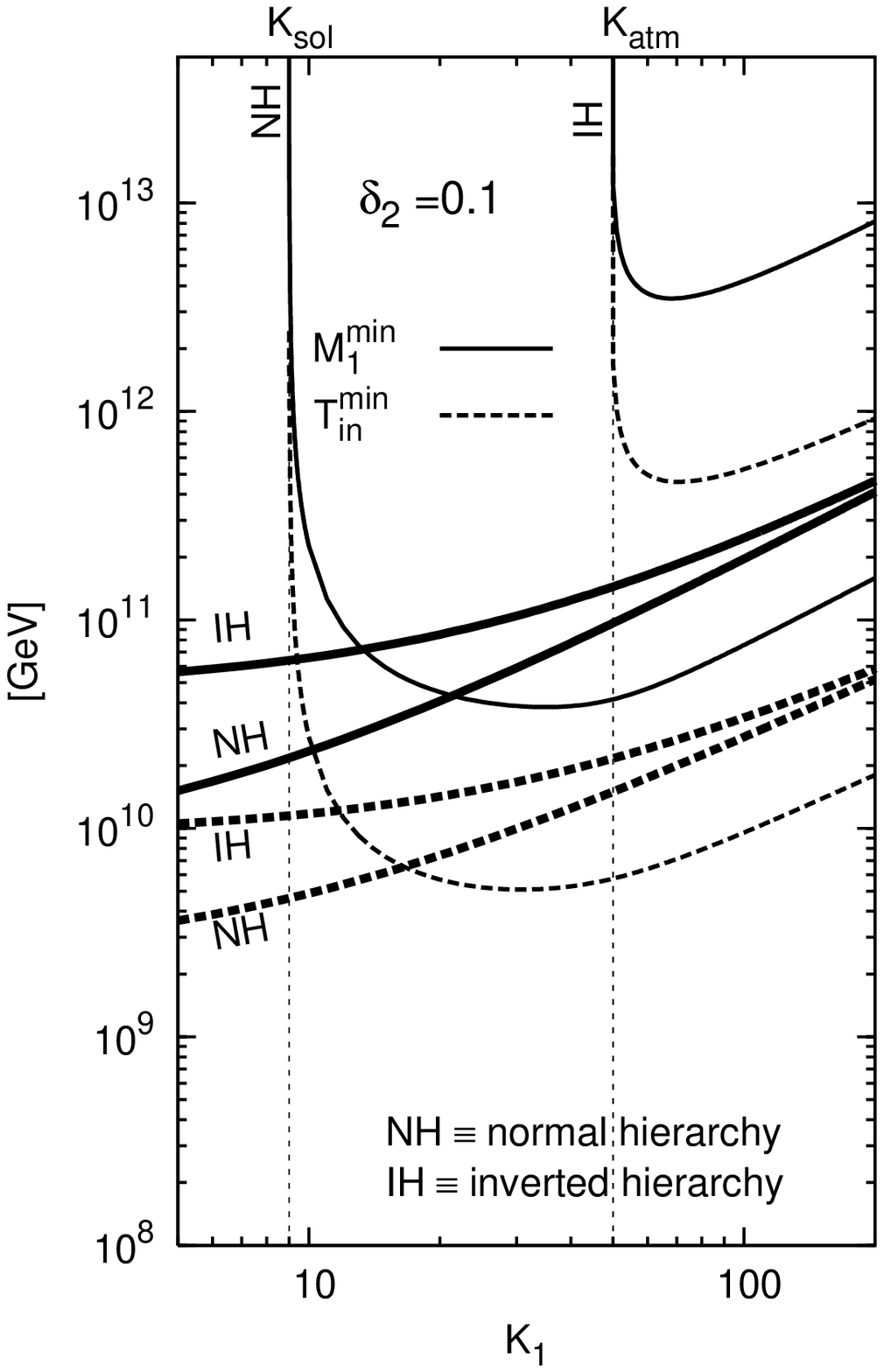,height=85mm,width=75mm}}
\centerline{\psfig{file=d201.eps,height=85mm,width=75mm} \hspace{-5mm}
\psfig{file=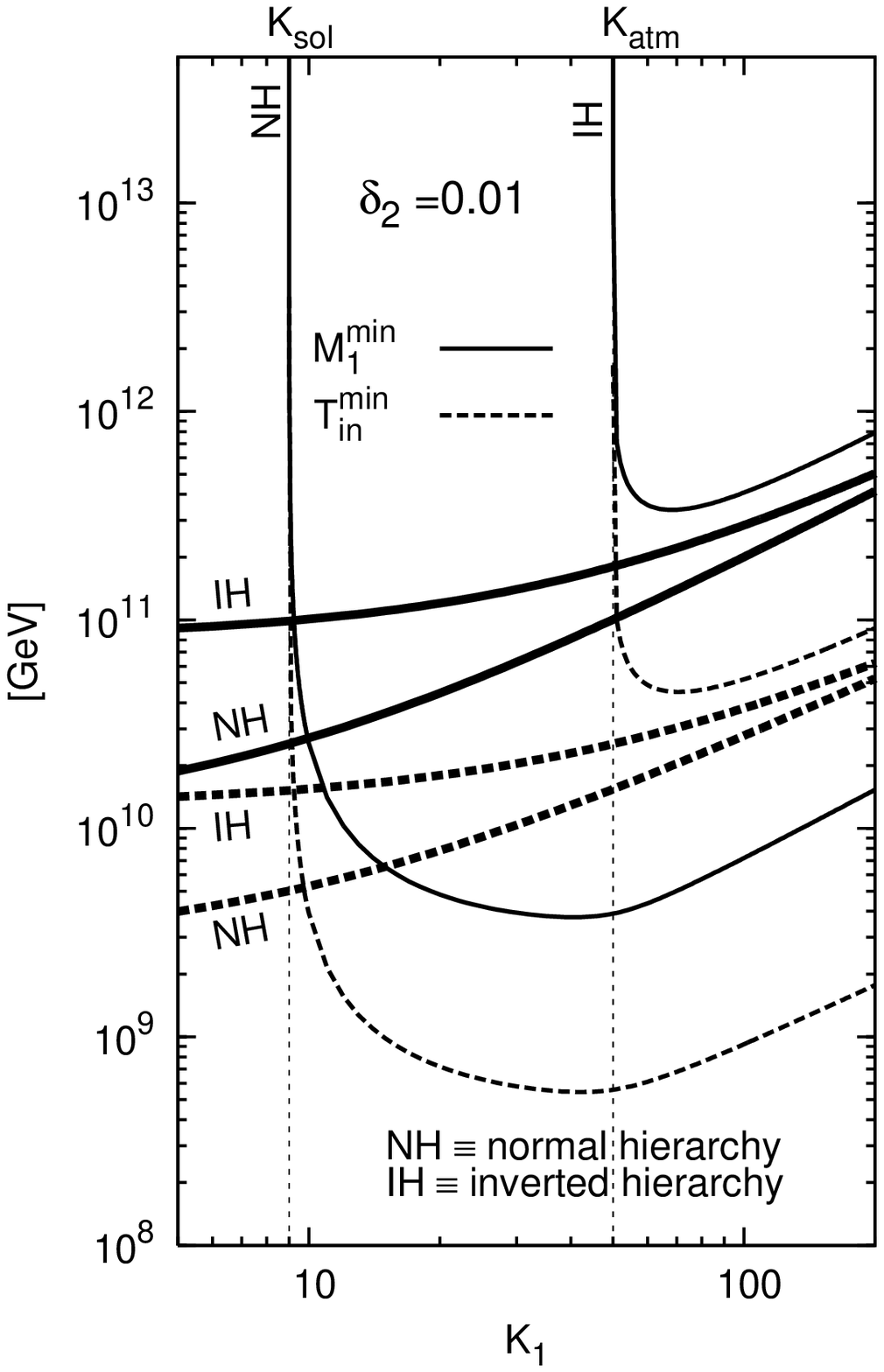,height=85mm,width=75mm}}
\caption{Lower bounds on $M_1$ and $T_{\rm in}$
in the case of `maximal phase' (thick lines)
and very large $M_3$ (thin lines). Upper panels: comparison
between the HL ($\d_2 \gg \d_2^{\rm HL}$) and $\d_2=0.1$.
Lower panels: comparison between $\d_2=0.1$ and $\d_2=0.01$.}
\end{figure}
The lower bounds for the very heavy $N_3$ case are indicated with
thin lines. One can see how these are relaxed just by a factor $2$
instead of one order of magnitude as one could expect from $C\!P$
asymmetry enhancement plus, sub-dominantly, the contribution to the final
asymmetry from $N_2$ decays. However, if one considers the DL, then
the asymmetry and the wash-out cumulative effects saturate,
while the $C\!P$ asymmetry enhancement
lowers the  bounds $\propto \xi(x_2)\simeq (3\,\d_2)^{-1}$
and this can be clearly described analytically. Indeed,
as it can be seen from Fig. 1, for small $\d_2$ one has
$\xi(1/x_2)=-\xi(x_2)$ and therefore, in the DL,
the expression for the final asymmetry simplifies
to (cf. (\ref{NBmLfss2}),(\ref{sind1}) and (\ref{k}))
\be
(N_{B-L}^{\rm f})_{\rm DL}(M_1,K_1,K_2)\equiv
2\,(1-\s^2)\,K_{\rm atm}\,\bar{\ve}(M_1)\xi(x_2)
\,z_B(K_1+K_2)\,{Y_{31}(K_1,K_2)\over K_1\,K_2} \propto \xi(x_2) \, .
\ee
In the bottom panel of Fig. 8 we compared
the lower bounds for $\d_2=0.1$ with those for $\d_2=0.01$
and one can see how the lower bound on $M_1$ ($T_{\rm in}$)
is $4\times 10^{10}\,{\rm GeV}$ ($5\times 10^{9}\,{\rm GeV}$)
for $\d_2=0.1$ and lowers to $4\times 10^{9}\,{\rm GeV}$
($5\times 10^{8}\,{\rm GeV}$) for $\d_2=0.01$. For
$\d_2\lesssim 0.01$, the wash-out and asymmetry cumulative effects
are negligible and the $C\!P$ asymmetry enhancement is the only left
one, so that
\be
(M_1^{\rm min})_{\rm DL} \simeq 4\times 10^9\,{\rm GeV}\,
\left({0.01\over \d_2}\right) \,
\hspace{9mm}
\mbox{and}
\hspace{9mm}
(T_{\rm in}^{\rm min})_{\rm DL} \simeq 5\times 10^8\,{\rm GeV}\,
\left({0.01\over \d_2}\right) .
\ee
The enhancement at small $\d_2$ has an upper limit set by the
condition of validity of the Eq.~(\ref{CPas}), which implies that
values $|\ve_1|\sim {\cal O}(0.01)$ can always be reached and thus that
there is no lower bound on $M_1$ if arbitrarily small $\d_2$ values
are possible. Examples of models where such small values can be
motivated are soft leptogenesis \cite{soft} and
radiative leptogenesis \cite{branco}.

In conclusion a strong relaxation of the lower bounds on $M_1$ and $T_{\rm in}$ is
attainable only for degeneracies $\d_2\ll 0.1$. This
confirms and actually strengthens the conclusions of \cite{bound2},
showing at the same time that effects arising from a term $\D\ve_1$ are
possible only for a very special category of models where a complex rotation $R_{23}$
is switched on with very high values of $|\O_{22}|$ implying
unlikely phase cancellations. This term can nevertheless play an interesting
role in relaxing the upper bound on the
absolute neutrino mass scale \cite{hambye}.

\section{Final Discussion}

We have calculated the final baryon asymmetry
in leptogenesis scenarios beyond the HL. Our main result
is that the lower bounds on $M_1$ and $T_{\rm in}$
can be relaxed but not necessarily, depending
on the particular form of the see-saw orthogonal matrix
and on the value of $\d_2$.
Very small values $\d_2\ll 0.1$ are  necessary
to have a significant relaxation of the lower bounds, confirming
and strengthening what was found in \cite{bound2}. One should then
understand, on theoretical grounds, whether such small values of $\d_2$
should be considered natural or not.

There is no compelling reason to exclude $M_1$ as large as
$10^{11}\,{\rm GeV}$ and $T_{\rm in}$ as large as
$10^{10}\,{\rm GeV}$. In many specific models
the lower bound on $M_1$ is however a problem,
especially in those cases where
the Dirac neutrino mass matrix is required to resemble the up
quark mass matrix \cite{kingsmirnov}.
In supergravity the gravitino problem requires
reheating temperatures that can be quite difficultly compatible
with the lower bound from leptogenesis in the HL.
In this case $\d_2\sim 10^{-3}$ or smaller would be desirable.
Within a minimal model with 3 RH neutrinos, the $N_2$DS
is an alternative solution.
Here one does not need to go beyond the HL for $M_1$
to be arbitrarily low, but
$T_{\rm in}\gtrsim 5\times 10^{9}\,{\rm GeV}$ still applies.
At the moment there is no reason to prefer one of the two
possible ways to go beyond the $N_1$DS, in any case
still the most appealing scenario. As we have seen,
a $N_2$DS is not possible for very large $M_3$, since in this
limit and for large $\d_2\gg 1$ the $C\!P$ asymmetry $\ve_2$
vanishes. It is important to notice moreover that the $N_2$DS
will be disproved if the absolute neutrino mass scale is found to be
$m_1\gtrsim (10^{-3}-10^{-2})\,{\rm eV}$.
On the other hand, we do not have currently any experimental test
to disprove small $\d_2$ values and there is a large freedom in the
choice of the heavy neutrino mass spectrum, also on pure
theoretical grounds.

It is quite interesting that a normal hierarchical
light neutrino spectrum is desirable since
an inverted one exacerbates the problem
 with the $M_1$ and $T_{\rm in}$ lower bounds.

 Notice that flavor effects \cite{bcst,newflavor1,newflavor2}
 introduce new interesting modifications of the lower bounds
but they do not change any of the conclusions we have
discussed here and will be addressed in a forthcoming paper.

A test of leptogenesis in the next years represents a great challenge.
However, in this respect, it is  more convenient to regard it
within the wider context of a test of the see-saw mechanism,
where leptogenesis is one among various phenomenological consequences.

A smoking gun seems possible
only in the very fortunate case that the lightest or the two lightest heavy
neutrinos are at the TeV scale and possess extra gauge interactions,
making possible their production and detection at the LHC. Because
the lower bound on $M_1$ is so large, this cannot happen in the HL
within the usual $N_1$DS but it becomes possible either in the DL with
$\d_2\lesssim 10^{-7}$ \cite{ky} or, more appealingly, in the $N_2$DS
\cite{geometry}. We can thus conclude that exciting results are
likely to come again in leptogenesis in the near future.

\vspace{2mm}
\noindent
\textbf{Acknowledgments}\\
It is a pleasure to thank A.~Anisimov, A.~Broncano, G.~Branco,
R.~Gonzalez Felipe, S.~Petcov, M.~Pl\"{u}macher,
G.~Raffelt and A.~Weiler for useful discussions.
\noindent



\begin{thebibliography}{99}

\bibitem{fy}
M.~Fukugita, T.~Yanagida, \pl{174}{1986}{45} \, .

\bibitem{seesaw}
P.~Minkowski, Phys.\ Lett.\ B {\bf 67} (1977) 421;
T.~Yanagida, in {\it{Workshop on Unified Theories}}, KEK report
79-18 (1979) p.~95;
M.~Gell-Mann, P.~Ramond, R.~Slansky, in {\it{Supergravity}} (North Holland,
Amsterdam, 1979) eds. P.~van Nieuwenhuizen, D.~Freedman, p.~315;
S.L. Glashow, in {\it 1979 Cargese Summer Institute on Quarks and Leptons}
(Plenum Press, New York, 1980) 
p.~687;
R.~Barbieri, D.~V.~Nanopoulos, G.~Morchio and F.~Strocchi,
Phys.\ Lett.\ B {\bf 90} (1980) 91;
R.~N.~Mohapatra and G.~Senjanovic,
Phys.\ Rev.\ Lett.\  {\bf 44} (1980) 912.

\bibitem{kt}
E.~W.~Kolb and M.~S.~Turner,
Ann.\ Rev.\ Nucl.\ Part.\ Sci.\  {\bf 33} (1983) 645.



\bibitem{annals}
W.~Buchm\"uller, P.~Di Bari and M.~Pl\"{u}macher,
Annals Phys.\  {\bf 315} (2005) 305.


\bibitem{asaka}
T.~Asaka, K.~Hamaguchi, M.~Kawasaki and T.~Yanagida,
Phys.\ Lett.\ B {\bf 464} (1999) 12.

\bibitem{di02}
S.~Davidson, A.~Ibarra, \pl{535}{2002}{25}.


\bibitem{cmb}
W.~Buchm\"{u}ller, P.~Di Bari and M.~Pl\"{u}macher,
  Nucl.\ Phys.\ B {\bf 643} (2002) 367.


\bibitem{geometry}
  P.~Di Bari, Nucl.\ Phys.\ B {\bf 727} (2005) 318.


\bibitem{giudice}
  G.~F.~Giudice, A.~Notari, M.~Raidal, A.~Riotto and A.~Strumia,
  Nucl.\ Phys.\ B {\bf 685} (2004) 89.


\bibitem{proc}
P.~Di Bari,
arXiv:hep-ph/0406115.


\bibitem{gkr}
G.~F.~Giudice, E.~W.~Kolb and A.~Riotto,
Phys.\ Rev.\ D {\bf 64} (2001) 023508.



%
%


\bibitem{bound1}
W.~Buchm\"{u}ller, P.~Di Bari and M.~Pl\"{u}macher,
Phys.\ Lett.\ B {\bf 547} (2002) 128.

\bibitem{bound2}
  W.~Buchm\"{u}ller, P.~Di Bari and M.~Pl\"{u}macher,
  Nucl.\ Phys.\ B {\bf 665} (2003) 445.


\bibitem{ke}
M.~Pl\"umacher, Z.~Phys.~{\bf C\ 74} (1997) 549.


\bibitem{many}
G.~C.~Branco, R.~Gonzalez Felipe, F.~R.~Joaquim,
I.~Masina, M.~N.~Rebelo and C.~A.~Savoy,
Phys.\ Rev.\ D {\bf 67} (2003) 073025;
A.~Pilaftsis and T.~E.~J.~Underwood,
Nucl.\ Phys.\ B {\bf 692} (2004) 303;
T.~Hambye, Y.~Lin, A.~Notari, M.~Papucci and A.~Strumia,
Nucl.\ Phys.\ B {\bf 695} (2004) 169.

\bibitem{branco}
 G.~C.~Branco, R.~Gonzalez Felipe, F.~R.~Joaquim and B.~M.~Nobre,
  Phys.\ Lett.\ B {\bf 633} (2006) 336.



\bibitem{hambye}
T.~Hambye, Y.~Lin, A.~Notari, M.~Papucci and A.~Strumia,
Nucl.\ Phys.\ B {\bf 695} (2004) 169.


\bibitem{bcst}
R.~Barbieri, P.~Creminelli, A.~Strumia and N.~Tetradis,
Nucl.\ Phys.\ B {\bf 575} (2000) 61.

\bibitem{newflavor1}
T.~Endoh, T.~Morozumi and Z.~h.~Xiong,
  Prog.\ Theor.\ Phys.\  {\bf 111} (2004) 123;
 E.~Nardi, Y.~Nir, E.~Roulet and J.~Racker,
  JHEP {\bf 0601} (2006) 164.

\bibitem{newflavor2}
A.~Abada, S.~Davidson, F.~X.~Josse-Michaux, M.~Losada and A.~Riotto,
  arXiv:hep-ph/0601083.

\bibitem{fgy}
P.~H.~Frampton, S.~L.~Glashow and T.~Yanagida,
Phys.\ Lett.\ B {\bf 548} (2002) 119.

\bibitem{luty}
  M.~A.~Luty,
  Phys.\ Rev.\ D {\bf 45} (1992) 455.

\bibitem{dolgov}
A.D.~Dolgov, Sov. J. Nucl. Phys. {\bf 32}, 831 (1980);
E.~W.~Kolb, S.~Wolfram, \np{172}{1980}{224}.

\bibitem{alicia}
W.~Buchm\"uller, M.~Pl\"umacher, \pl{431}{1998}{354};
A.~Anisimov, A.~Broncano and M.~Pl\"{u}macher, Nucl.\ Phys.\ B {\bf 737} (2006) 176.

\bibitem{CPas}
M.~Flanz, E.~A.~Paschos, U.~Sarkar, \pl{345}{1995}{248};
\pl{384}{1996}{487} (E)
L.~Covi, E.~Roulet, F.~Vissani, \pl{384}{1996}{169}


\bibitem{casas}
J.~A.~Casas and A.~Ibarra,
Nucl.\ Phys.\ B {\bf 618} (2001) 171.

\bibitem{WMAPSLOAN}
WMAP Collaboration, D.~N.~Spergel {\it et al.},
Astrophys.\ J.\ Suppl.\  {\bf 148} (2003) 175;
M.~Tegmark {\it et al.}  [SDSS Collaboration],
 Phys.\ Rev.\ D {\bf 69} (2004) 103501.


\bibitem{vives}
O.~Vives, arXiv:hep-ph/0512160.

\bibitem{ir}
A.~Ibarra and G.~G.~Ross, Phys.\ Lett.\ B {\bf 575} (2003) 279.

\bibitem{ct}
P.~H.~Chankowski and K.~Turzynski, Phys.\ Lett.\ B {\bf 570} (2003) 198.


\bibitem{soft}
  G.~D'Ambrosio, G.~F.~Giudice and M.~Raidal,
  Phys.\ Lett.\ B {\bf 575} (2003) 75.

\bibitem{kingsmirnov}
G.~C.~Branco, R.~Gonzalez Felipe, F.~R.~Joaquim and M.~N.~Rebelo,
Nucl.\ Phys.\ B {\bf 640} (2002) 202;
S.~F.~King, Rept.\ Prog.\ Phys.\  {\bf 67} (2004) 107;
E.~K.~Akhmedov, M.~Frigerio and A.~Y.~Smirnov, JHEP {\bf 0309} (2003) 021.


\bibitem{ky}
 A.~Pilaftsis,
  Phys.\ Rev.\ D {\bf 56} (1997) 5431;
  S.~F.~King and T.~Yanagida,
  Prog.\ Theor.\ Phys.\  {\bf 114} (2006) 1035.


\end{thebibliography}
\end{document}